\documentclass[sigconf]{acmart}
\graphicspath{{imgs/}}
\usepackage{etex}
\usepackage{microtype}
\usepackage{xcolor}
\usepackage{longtable}
\usepackage{balance}
\usepackage{url}
\usepackage{graphicx}
\usepackage{array}
\usepackage{amsmath,mathtools}
\usepackage[ruled,vlined,linesnumbered]{algorithm2e}
\usepackage{verbatim}
\usepackage{comment}
\usepackage[normalem]{ulem}
\usepackage{enumitem}
\usepackage{multicol}
\usepackage{epsfig}
\usepackage{endnotes}
\usepackage{subfig}
\usepackage{pgfplots,pgfplotstable} 
\pgfplotsset{compat=newest}
\usepackage[T1]{fontenc}
\def\Snospace~{\S{}}
\def\Fnospace~{\mbox{Figure\hspace{0.25em}}}
\def\Tnospace~{\mbox{Table\hspace{0.25em}}}
\def\Enospace~{\mbox{Equation\hspace{0.25em}}}
\def\SEnospace~{\mbox{Section\hspace{0.25em}}}

\newcommand{\tinyskip}{\vspace{1pt}}
\newcommand{\mypar}[1]{\tinyskip\tinyskip\noindent\textbf{#1.}\xspace}
\newcommand{\myparr}[1]{\tinyskip\tinyskip\noindent\textbf{#1}\xspace}

\newcommand{\vspacebfigure}{\vspace{-6mm}}
\newcommand{\vspaceafigure}{\vspace{-2mm}}

\definecolor{lightgray}{rgb}{.85,.85,.85}  %
\definecolor{orange}{RGB}{255,127,0}
\newcommand{\mynote}[1]{\par\noindent\colorbox{lightgray}{\parbox{\linewidth}{#1}}}

\newcommand{\KK}[1]{\textcolor{blue}{\mynote{KK:~#1}}}
\newcommand{\CC}[1]{\textcolor{red}{\mynote{CC:~#1}}}
\newcommand{\YZ}[1]{\textcolor{magenta}{\mynote{YZ:~#1}}}
  \renewcommand{\KK}[1]{\null}
  \renewcommand{\YZ}[1]{\null}
  \renewcommand{\CC}[1]{\null}

\newcommand{\eat}[1]{}

\def\compactify{\noitemsep \itemsep=0pt \topsep=0pt \partopsep=0pt \parsep=0pt}
\let\latexusecounter=\usecounter

\newcommand{\sysname}{KEA\xspace}
\newcommand{\kea}{\sysname}
\newcommand{\tone}{Observational Tuning\xspace}
\newcommand{\ttwo}{Experimental Tuning\xspace}
\newcommand{\tthree}{Hypothetical Tuning\xspace}
\newcommand{\mysim}{\raise.17ex\hbox{$\scriptstyle\sim$}}

\AtBeginDocument{%
  \providecommand\BibTeX{{%
    \normalfont B\kern-0.5em{\scshape i\kern-0.25em b}\kern-0.8em\TeX}}}

\newcounter{enum}

\copyrightyear{2021} 
\acmYear{2021} 
\setcopyright{acmcopyright}\acmConference[SIGMOD '21]{Proceedings of the 2021 International Conference on Management of Data}{June 20--25, 2021}{Virtual Event, China}
\acmBooktitle{Proceedings of the 2021 International Conference on Management of Data (SIGMOD '21), June 20--25, 2021, Virtual Event, China}
\acmPrice{15.00}
\acmDOI{10.1145/3448016.3457569}
\acmISBN{978-1-4503-8343-1/21/06}

\begin{document}
\fancyhead{}
\title{\kea: Tuning an Exabyte-Scale Data Infrastructure}

\newcommand{\autspace}{~~~~~~~~~~}
\author{
	Yiwen Zhu$^1$ \autspace Subru Krishnan$^1$ \autspace Konstantinos Karanasos$^1$ \autspace Isha Tarte$^1$ \autspace Conor Power$^1$  \autspace Abhishek Modi$^1$
	\and
	Manoj Kumar$^1$ \autspace Deli Zhang$^1$ \autspace Kartheek Muthyala$^1$ \autspace Nick Jurgens$^1$ \autspace Sarvesh Sakalanaga$^2$}
\authornote{The work was done while the author was at Microsoft.}
\author{
	Sudhir Darbha$^1$  \autspace Minu Iyer$^1$ \autspace Ankita Agarwal$^1$ \autspace Carlo Curino$^1$
	\and
}

\affiliation{ 
	\institution{$^1$Microsoft, firstname.lastname@microsoft.com}
}
\affiliation{ 
	\institution{$^2$Salesforce, firstname.lastname@salesforce.com}
}

\renewcommand{\shortauthors}{Zhu et al.}
\begin{abstract}
Microsoft's internal big-data infrastructure is one of the largest in the world---with over 300k machines running billions of tasks from over 0.6M daily jobs. Operating this infrastructure is a costly and complex endeavor, and efficiency is paramount. In fact, for over 15 years, a dedicated engineering team has tuned almost every aspect of this infrastructure, achieving state-of-the-art efficiency (>60\% average CPU utilization across all clusters).  Despite rich telemetry and strong expertise, faced with evolving hardware/software/workloads this manual tuning approach had reached its limit---we had plateaued. 

In this paper, we present \sysname, a multi-year effort to automate our tuning processes to be fully data/model-driven. \sysname leverages a mix of domain knowledge and principled data science to capture the essence of our cluster dynamic behavior in a set of machine learning (ML) models based on collected system data. These models power automated optimization procedures for parameter tuning, and inform our leadership in critical decisions around engineering and capacity management (such as hardware and data center design, software investments, etc.). We combine ``observational'' tuning (i.e., using models to predict system behavior without direct experimentation) with judicious use of ``flighting'' (i.e., conservative testing in production). This allows us to support a broad range of applications that we discuss in this paper.  

\sysname continuously tunes our cluster configurations and is on track to save Microsoft tens of millions of dollars per year. At the best of our knowledge, this paper is the first to discuss research challenges and practical learnings that emerge when tuning an exabyte-scale data infrastructure.

\end{abstract}

\maketitle

\section{Introduction}
\label{sec:intro}

Big-data infrastructures have empowered users with unprecedented ability to store and process large amounts of data, paving the road for revolutions in the areas of web search, analytics, and artificial intelligence~\cite{zaharia2010spark, thusoo2009hive,tigani2014google,chaiken2008scope,ramakrishnan2017azure,aws-athena, sarkar2018learning,amershi2019software,davenport2018using}.

\begin{figure}[!t]
	\centering
	\includegraphics[width=\columnwidth]{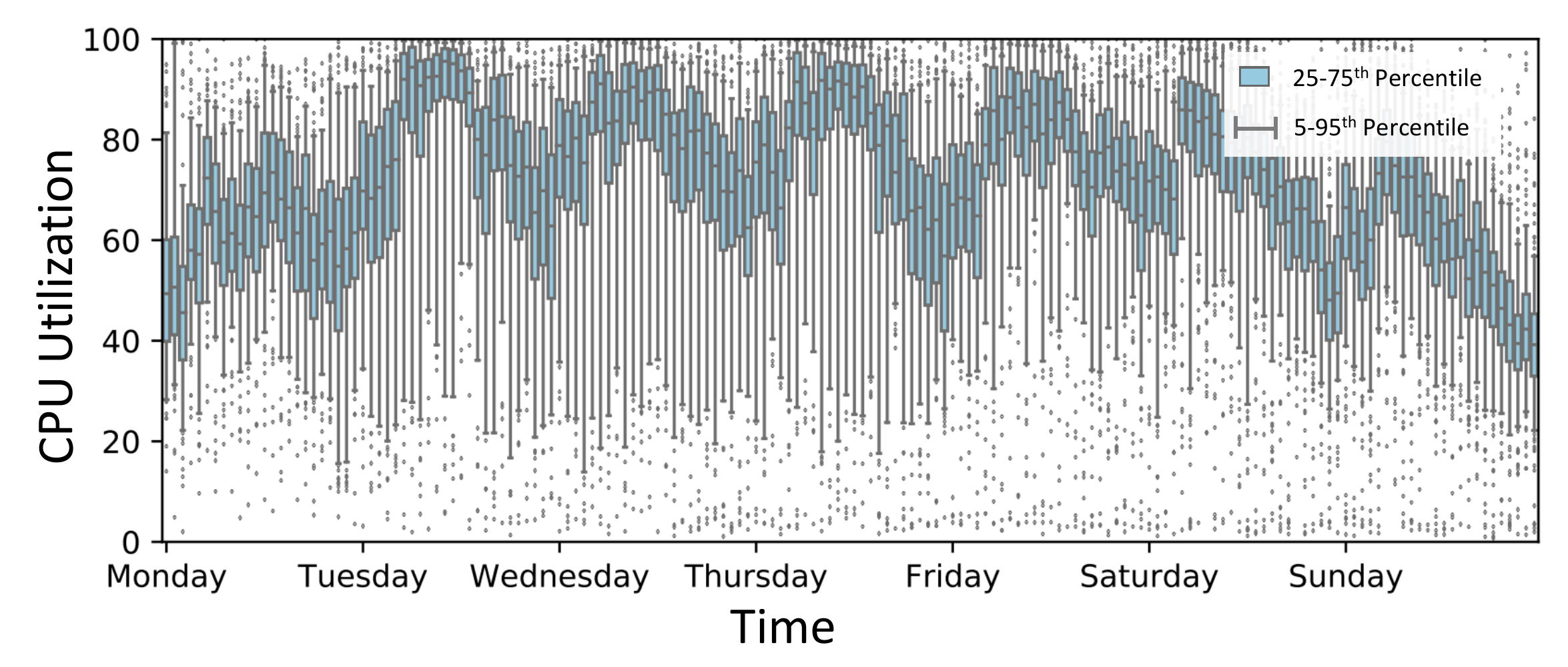}
	\vspacebfigure
	\vspace{-0.1cm}
	\caption{CPU utilization for a typical week}\label{fig:cpu}
	\vspaceafigure
	\vspace{-0.2cm}
\end{figure}

At Microsoft, we operate \textit{Cosmos}, one of the largest big-data infrastructures in the world, with over 300k machines serving billions of tasks daily. Tremendous engineering and research effort has been devoted over the last 15 years to improve Cosmos' scalability, efficiency, security, and reliability~\cite{jindal2019peregrine, zhang2012optimizing, bruno2012recurring, bruno2013continuous,boutin2014apollo,boutin2015jetscope,karanasos2015mercury,jyothi2016morpheus,curino2019hydra,qiao2019hyper,chaiken2008scope,ferguson2012jockey,ramakrishnan2017azure,zhou2012advanced,zhou2010incorporating,wing2020osdi}.
Given the significance and complexity of this infrastructure, almost every aspect of it has been carefully tuned by a dedicated team to optimize its efficiency, improving its performance and reducing operational costs.
As a result, the team has reached industry-leading levels of utilization (e.g., >60\% average CPU utilization, as shown in Figure~\ref{fig:cpu}). 
But despite the domain expertise and rich telemetry, our improvements started to plateau as we reached the limits of what we could achieve via manual tuning.

To overcome this plateau, we turned to very large-scale data science as a tool to further optimize our operations.
Although such ``ML-for-Systems'' approaches have been widely employed in multiple settings  lately~\cite{curino2020mlos}, unsurprisingly, 
they are impractical at our scale. 
Among them, the more practically viable ones have focused on small-scale settings (e.g., to tune a single DBMS instance), being based on the fundamental assumption of repeatedly changing system tunables and running experiments to measure the resulting performance~\cite{alipourfard2017cherrypick,kanellis2020too,van2017automatic,pavlo2019external,johannesBW}.
This \emph{``experimental tuning''} approach can be unrealistic in our setting: deployments must roll out progressively across tens of thousands of machines, noisy workloads require long windows of observation (>weeks), and running a ``bad'' configuration could have devastating effects on some of the most business-critical operations at Microsoft. As we will show, we use experimental tuning only as a last resort.

The system and methodology we developed, named \emph{\sysname}, is the result of a multi-year effort aimed at evolving our processes to be fully automated and data/model-driven. \sysname systematically combines domain expertise and principled data science to capture the complex dynamic behavior of a big-data cluster as a collection of ``descriptive'' and ``predictive'' ML models, i.e., models that can describe and predict the system's behavior. These models power automated optimization procedures and are used both for direct parameter tuning, but also to guide our leadership in tactical engineering and capacity decisions (such as hardware and data center design, software investments, etc.). \sysname employs rich ``observational'' tuning (i.e., without requiring to modify the system) with judicious use of ``flighting'' (i.e., conservative testing in production). This allows us to support a broad range of tuning applications.

We suspect that a handful of other companies are operating solutions akin to what we describe here, but to the best of our knowledge this is the first paper describing the data science-driven tuning of an exabyte-scale analytics infrastructure. In presenting this work, we focus on the peculiar challenges that arise at scale, and present a methodology and tools to tackle them, as well as lessons learned in deploying these ideas in production settings.

In summary, we make the following contributions:
\begin{itemize}
	\item 
	We define a methodology to cope with the system complexity and create compact, sound, and explainable models of a cloud infrastructure based on a set of tractable metrics, which can explain ``why'' the optimal configuration is chosen. 
	\item We present the end-to-end architecture of \sysname and provide details for the three types of tuning that \sysname enables: (i)~\emph{observational} tuning, which employs models for picking the right parameters, avoiding costly rounds of experiments; (ii)~\emph{hypothetical} tuning, an ML-assisted methodology for future planning; and (iii)~\emph{experimental} tuning, our fall-back approach that judiciously performs experiments when it is not possible to predict the system behavior otherwise.
	\item Deployed in production, \sysname continuously tunes our Cosmos clusters and is on track to save the company tens of millions of dollars yearly. We share our experience of running \sysname in production at such a massive scale. 
\end{itemize}

The remainder of this paper is organized as follows. 
Section~\ref{sec:bg} provides background and motivates our cluster tuning problem.
Section~\ref{sec:rcp} presents the rationale behind our system conceptualization and the derivation of relevant metrics. 
Section~\ref{sec:overview} describes the overall architecture and summarizes the different tuning modes.
Sections~\ref{sec:tuning},~\ref{sec:skuDesign}, and ~\ref{sec:pc} describe each tuning approach in more detail. 
Section~\ref{sec:related} discusses related work, then we conclude.

\begin{table}[t]
	\caption{Cosmos statistics~\cite{curino2019hydra,shao2019griffon}}\label{tab:cosmos}
	\vspaceafigure
	\begin{tabular}{ll}
		\toprule
		Description & Size                              \\ \midrule
		Number of jobs per day     & >600k             \\
		Number of tasks per day     & >4B             \\
		Number of users   & >10k\\
		Total number of machines      & >300k        \\
		Number of machines per cluster         & >45k         \\
		Total Hardware Capital Expenditure in US Dollar    & >\$1B             \\
		\bottomrule
	\end{tabular}
		\vspace{-0.1cm}
\end{table}

\section{Background \& Motivation}\label{sec:bg}

At Microsoft, we operate \textit{Cosmos}, a data analytics platform for the internal analytics needs across the company~\cite{curino2019hydra}. 
Cosmos is one of the largest data infrastructures worldwide with more than 300k machines spread across several data centers.
Cluster resources are spread across tens of thousands of users that submit more than half a million analytics jobs daily. 
\autoref{tab:cosmos} summarizes several scalability metrics of this massive infrastructure.

\begin{figure}[t]
	\centering
	\begin{minipage}[b]{0.47\columnwidth}
	\includegraphics[width=\textwidth]{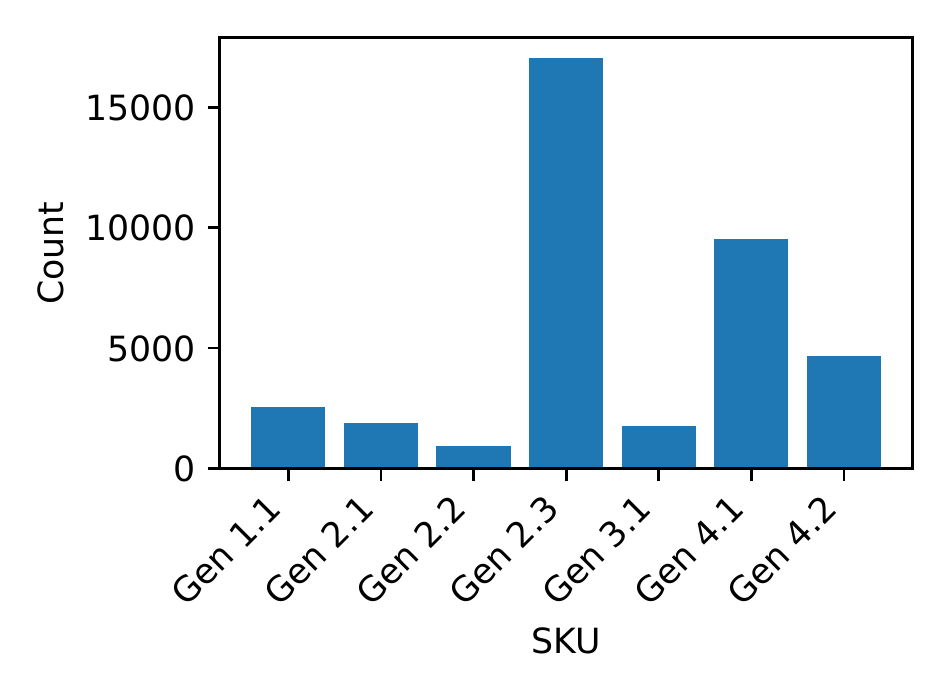}
	\end{minipage}%
	\begin{minipage}[b]{0.53\columnwidth}
	\includegraphics[width=\textwidth]{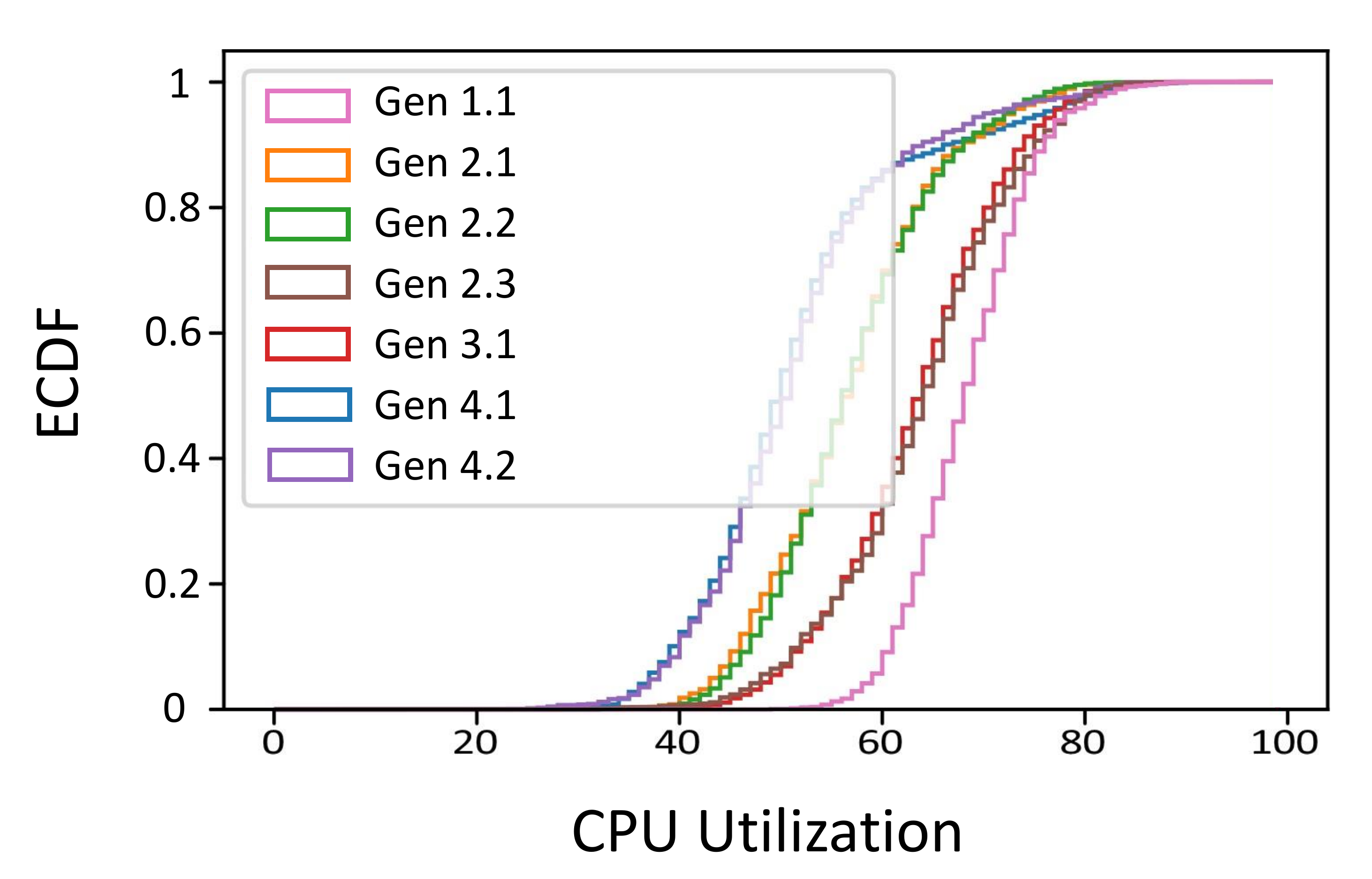}
	\end{minipage}%
	\vspace{-2mm}
	\caption{Machine count (left) and utilization level (right) for different hardware generations for one of the clusters}\label{fig:count0} 
	\vspaceafigure\vspace{-2mm}
\end{figure}
The vast majority of the submitted jobs are written in Scope~\cite{zhou2012scope}, a SQL-like dialect (with heavy use of C\# and Python UDFs). 
Scope jobs are translated to a DAG of operators that are spread for execution across several machines. Each job is comprised of up to hundreds of thousands of tasks, i.e., individual processes each executed in one \textit{container}.
A YARN-based~\cite{vavilapalli2013apache} resource manager is used for scheduling tasks and sharing cluster resources across jobs~\cite{curino2019hydra}. 
More than four billion tasks per day get scheduled across Cosmos.

To make matters even more complicated, along with scale comes heterogeneity.
After a decade of operation, Cosmos involves more than 20 hardware generations of machines (with varying CPU cores, RAM, HDD/SSD) from various manufacturers and software configurations (e.g., mapping of drives to SSDs/HDDs). Hereafter, we use the term stock keeping unit, or \emph{SKU}, to refer to a hardware generation and \emph{SC} to a software configuration. Each cluster consists of 6 to 9 SKUs for working machines (compute nodes), which are the main focus of this work.
\autoref{fig:count0} shows the distribution of machines per SKU for one of the clusters.

Operating such a complex infrastructure requires tuning hundreds of parameters across user applications and the underlying infrastructure. In this paper, we focus on infrastructure-level configurations, and in particular on cluster-wide configurations, as they are very impactful and traditionally harder to tune. As we show, at this scale we cannot simply perform tuning based on a large round of experiments or trivial A/B testing---a more sophisticated approach is required.

At the same time, the starting point for our optimization journey is a non-trivial one to beat: years of domain experts fine-tuning Cosmos to maximize efficiency and reach an industry-leading CPU utilization of over 60\% (average across all our clusters).
However, this manual effort is time-consuming and error-prone, and it requires continued adjustments as workloads shift and new hardware and software is deployed.
Having no rigorous way to evaluate the current parameter choice and suggest more promising values leads to significant missed opportunities in terms of operational cost and performance.

Tuning cluster-wide parameters is a scary and time-consuming affair, as changes must be rolled-out progressively across the fleet, mistakes are costly as performance may crater, and long-term workload seasonalities impose long observation windows. This leads to very slow progress as operators need to proceed very cautiously and validate for long periods of time. A symptom of this becomes directly evident on the right-hand side of \autoref{fig:count0}: while overall utilization is good, older-generation machines (i.e., the ones we have been tweaking for longer) are substantially more utilized, as we have slowly learned how hard we can push the hardware before negatively impacting stability or user-perceived performance. As our workloads keep evolving and we continuously roll out new (and thus never-tested-before) SKUs, this slow time-to-optimal tuning puts a cap on our maximum efficiency (as a fraction of workloads/hardware is always sub-optimally used).

Realizing the limitations of the current manual approach in parameter tuning, three years ago we embarked in a journey to tune our clusters in a principled data-driven way, and the \kea project was born.

\section{End-to-end Methodology}\label{sec:rcp}

With \sysname we introduce both a methodology and a software architecture to support tuning of large scale systems. In this section, we focus on the methodological aspects, which are designed to guide a data science team to interact with the engineering group owning and operating the system to be tuned. The KEA software architecture is discussed in Section~\ref{sec:overview}.

\subsection{Phases of a KEA project}
\begin{figure}[!t]
	\centering
	\includegraphics[width=\columnwidth]{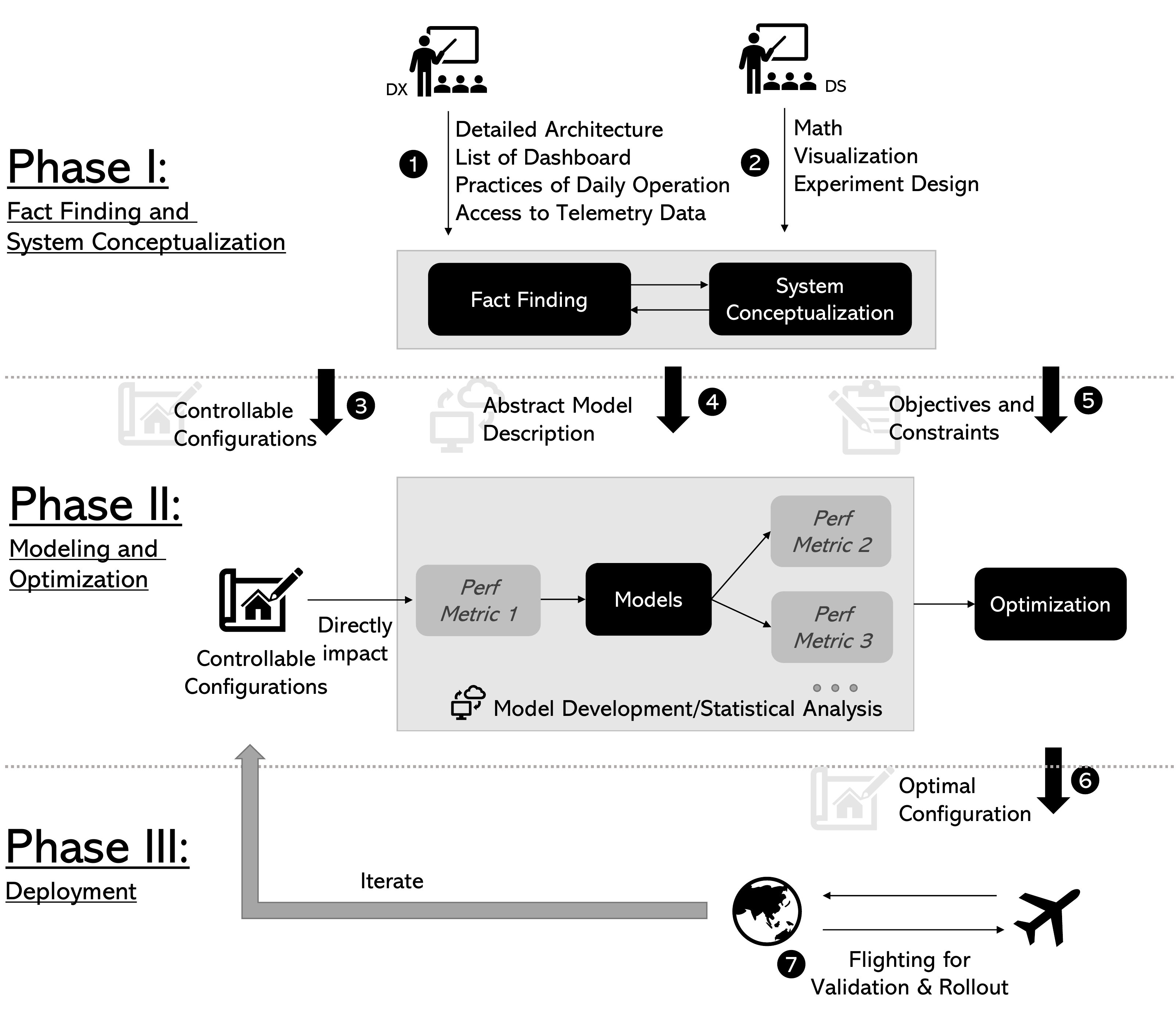}
	\vspace{-0.6cm}
	\caption{End-to-end methodology of KEA}\label{fig:rcp}
	\vspace{-0.3cm}
\end{figure}

The KEA methodology is structured in three distinct phases, as shown in Figure~\ref{fig:rcp}.

\noindent {\bf Fact Finding and System Conceptualization (Phase I).} In this phase, all the stakeholders in a ``tuning project'' come together to identify the goal and scope of the project. More precisely, the data scientists (DS) engage with the engineering team owning the system, referred hereafter as the domain experts (DX), in a series of interviews and brainstorming sessions aimed at \textit{Fact Finding} and \textit{System Conceptualization}. 
\textit{Fact finding} aims at obtaining a thorough architecture description of the system, collecting intuition on the system dynamics, defining the objectives of the project (e.g., improving latency/operational cost), and identifying the controllable configurations.
The DS are granted access to telemetry data and dashboards used by DX for daily monitoring of the clusters, and come to learn the various practices of daily operation of the system, such as workflows for debugging and incident resolution (which are typically rich with data-to-insight information).
\textit{System conceptualization} aims at simplifying the complex architecture of the analyzed system to a smaller number of blocks with annotations of flows that describe the interaction between them. This is crucial to capture the system complexity in a compact and mathematically tractable way. 
The DS and DX identify the potential practical constraints that need to be satisfied in the tuning process, such as ensuring latency requirements. 
Neither DS nor DX can achieve this step alone and a close collaboration is required to leverage the collective expertise and knowledge. We describe the system abstractions we employ during conceptualization in \autoref{sec:metrics}.

In this conceptualization effort, DS leverages data-driven and statistical tools to observe correlations between key performance metrics, and collaborate with DX to understand the causal relationships between components, creating a correct model of the complex system. The combined domain knowledge and numerical validation creates a deeper understanding of the system, and it is in its own right valuable to the system operators (often the resulting visualizations are embraced by the engineering teams, even beside the rest of our automation work). Note that at this stage we have not built ML models yet, just a crisp understanding of how components relate to each other and what we could expect to predict effectively.

The {\em output} of Phase I includes the list of controllable configurations, the description of possible models to build in Phase II, along with the objectives and constraints that need to be considered.

\noindent {\bf Modeling and Optimization (Phase II).} In this phase, we use the \emph{system conceptualization} from Phase I and dive into a more quantitative analysis with actual ML, optimization, statistical, and econometric methods to model the behavior of the system using mathematical formulations based on appropriate assumptions. The result is a model of the entire system we can use to answer interesting ``what if'' questions.
The optimizer uses this model to pick the optimal configurations.
This phase mostly involves DS for the analysis and model development, while results are interpreted and validated by DX.

The {\em output} of this phase includes the optimal configuration setting for the target parameter(s) that maximizes the chosen objective function, subject to the constraints we defined in Phase I.

\begin{figure*}[!t]
	\centering
	\vspace{-3mm}
	\includegraphics[width=0.97\textwidth]{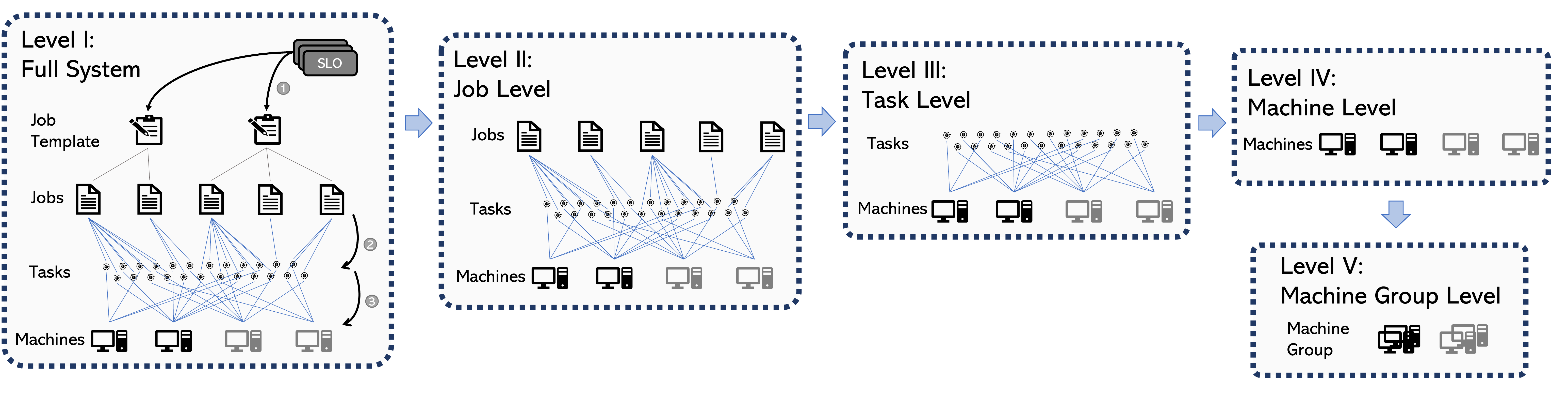}
	\vspace{-3mm}
	\caption{Simplifying Cosmos: from job-level SLO compliance to machine-centric metrics}\label{fig:metrics}
	\vspace{-2mm}
\end{figure*}

\noindent {\bf Deployment (Phase III)}. In this phase, we conduct ``flighting'' of the proposed configuration for validating the models and assumptions. The flighting service is an internal tool that allows cluster operators to deploy new software versions and new configuration parameters to a live production environment. The infrastructure gives us flexibility to deploy to partial or complete clusters, and to carefully measure the impact on user-observable and cluster-internal internal metrics (e.g., CPU utilization per node). Iteratively, DS fine-tunes the models and works closely with DX to monitor the cluster behavior and prepare for the final configuration roll-out.

\subsection{System Abstractions}\label{sec:metrics}

In this section, we discuss the abstractions we introduce to allow us to model the highly complex infrastructure we want to tune in a tractable manner. As discussed above, these abstractions play a key role in the \emph{system conceptualization} of Phase I.

Modeling all individual components of a large-scale infrastructure is intimidating as it involves a multitude of components, while detailed simulation or modeling of each aspect of the system is often intractable. 
For example, when applying \sysname to the Cosmos infrastructure, we must consider the behavior of and the interactions between the scheduler, compiler, query optimizer, job manager, and task executor. 
Predicting the runtime of hundreds of thousands of jobs by modeling the task-level interference across billions of tasks spanning tens of thousands of machines would require models with a massive number of parameters and unacceptable training times. 
Modeling abstractions and simplifications are crucial. 
In this paper, we demonstrate that the right abstraction process can lead to a handful of small models, which in turn are capable of capturing enough of a system behavior to inform the selection of an optimal configuration. As we show in Sections~\ref{sec:tuning}--\ref{sec:pc}, this process outperforms the existing manual tuning approach.

To make this modeling more concrete, we now discuss the introduced abstractions we made to model Cosmos. 
Our end goal is to improve Cosmos efficiency, i.e., pick new configurations that will allow us to run more jobs without affecting existing job runtime performance. 

{\color{black}\mypar{Full system abstraction (Level I)}} On the left of Figure~\ref{fig:metrics}, we depict the full complexity of Cosmos. Each machine runs a mix of tasks that belong to many job instances of various job ``templates''.\footnote{A job template represents a recurring job. It is created by removing the specific data inputs from the corresponding job script~\cite{curino2019hydra}.} Modeling the system by including all task-level interferences and intra-job dependencies is prohibitive. We thus describe several simplification steps (from left to right in the figure) leading to increasingly more manageable levels of abstraction, as shown in Figure~\ref{fig:metrics} and described below. {\color{black}Note that our abstractions take advantage of the fact that Cosmos uses a monolithic resource manager that assigns directly tasks to machines, and the majority of jobs are SCOPE jobs (see \autoref{sec:bg}). In the case of multi-framework two-level schedulers, such as Mesos~\cite{hindman2011mesos} or Borg~\cite{verma2015large}, similar abstractions could be made for different groups of machines (e.g., based on the jobs that typically land on each group).}

\mypar{Job-level abstraction (Level II)} The initial intuition on how to simplify this picture comes from the DX who informed the team that ``most jobs in Cosmos have implicit runtime SLOs''. This means that users expect a predictable behavior but don't have explicitly defined SLOs. In practice this means that recent runtime behaviors of a job induce an implicit SLO on the next execution of the same job template~\cite{jyothi2016morpheus,wing2020osdi}. This allows us to quantify a vague customer expectation in the following constraint: $\forall i,~ \text{runtime}(\text{job}_i,\text{conf}_{\text{new}}) \leq \text{runtime}(\text{job}_i, \text{conf}_{\text{old}})$, which states that the runtime of $\text{job}_i$ should be no worse with the new configuration ($\text{conf}_{\text{new}}$) than with the old configuration ($\text{conf}_{\text{old}}$).\footnote{The notation is simplified for the sake of presentation; more precisely we should compare the runtime of an upcoming job with an amortized historical runtime of instances from the same job template.} 
These constraints are statistical in nature due to naturally occurring variances based on job specification, input data, and transient cluster conditions. 
The DS quantifies the natural variance and leverages statistical testing techniques to verify whether the constraint holds during our experimentation. 

\eat{
\begin{figure}[t]
	\centering
	\begin{minipage}[b]{0.25\textwidth}
		\includegraphics[width=\textwidth]{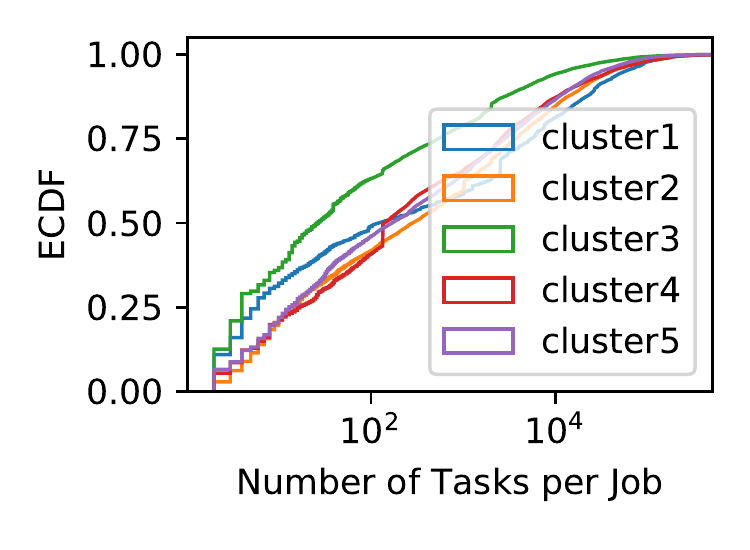}
	\end{minipage}%
	\begin{minipage}[b]{0.25\textwidth}
		\includegraphics[width=\textwidth]{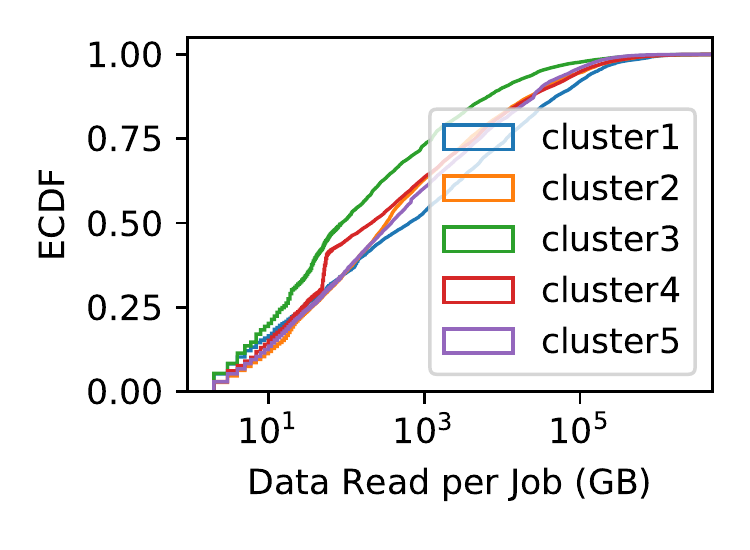}
	\end{minipage}%
	\caption{Task count (left) and Total Data Read (right) per job for different clusters. The variation across workloads is significant and it is hard to extract "representative workloads" and use them as canary jobs to take a peek at the system operation.}\label{fig:jobstats}
\end{figure} 
}

\begin{figure}[t]
	\centering
	\begin{minipage}[b]{0.5\columnwidth}
		\includegraphics[width=\textwidth]{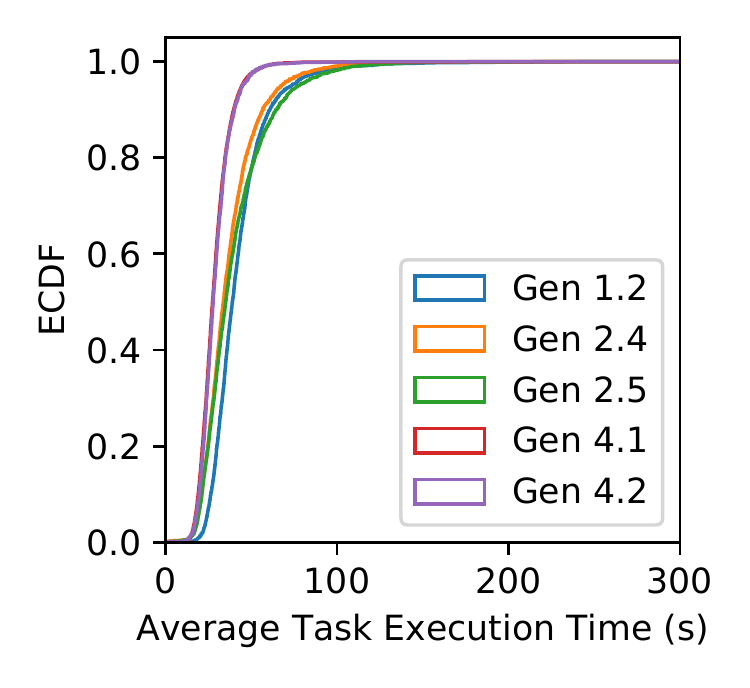}
	\end{minipage}%
	\begin{minipage}[b]{0.5\columnwidth}
		\includegraphics[width=\textwidth]{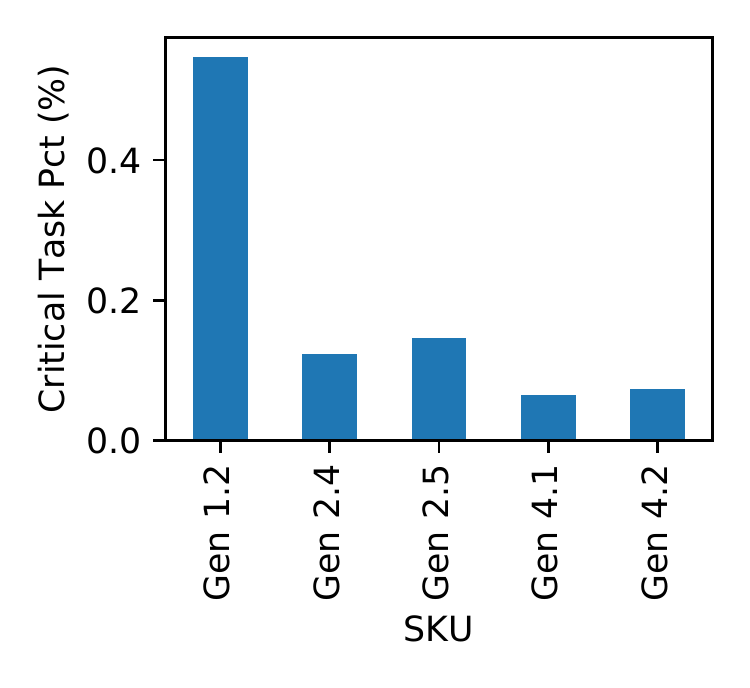}
	\end{minipage}%
	\vspace{-0.4cm}
	\caption{Task execution time distribution across different SKUs indicates that tasks executed on slower machines are more likely to be on the critical path of a job execution.}\label{fig:critical}
	\vspace{-0.2cm}
\end{figure}

\mypar{Task-level abstraction (Level III)} The next intuition from the DX is that ``job runtimes are dominated by slow tasks in the critical path'' (i.e., a set of slowest tasks in each stage of the job execution)~\cite{shao2019griffon}. The important implication is that ensuring that the performance of slow tasks is not impaired by a configuration change is a sufficient condition to satisfy the job-level performance constraint introduced above. 
The DS validates this by confirming that slow tasks are indeed on the critical path. 
The validation is more rich and complex, but the high level idea can be peeked in Figure~\ref{fig:critical}: tasks landing on slower machines (that are older and busier) are disproportionately more likely to be slower and therefore be part of the job's critical path.

\begin{figure}[t]
	\centering
	\begin{minipage}[b]{0.5\columnwidth}
		\includegraphics[width=\textwidth]{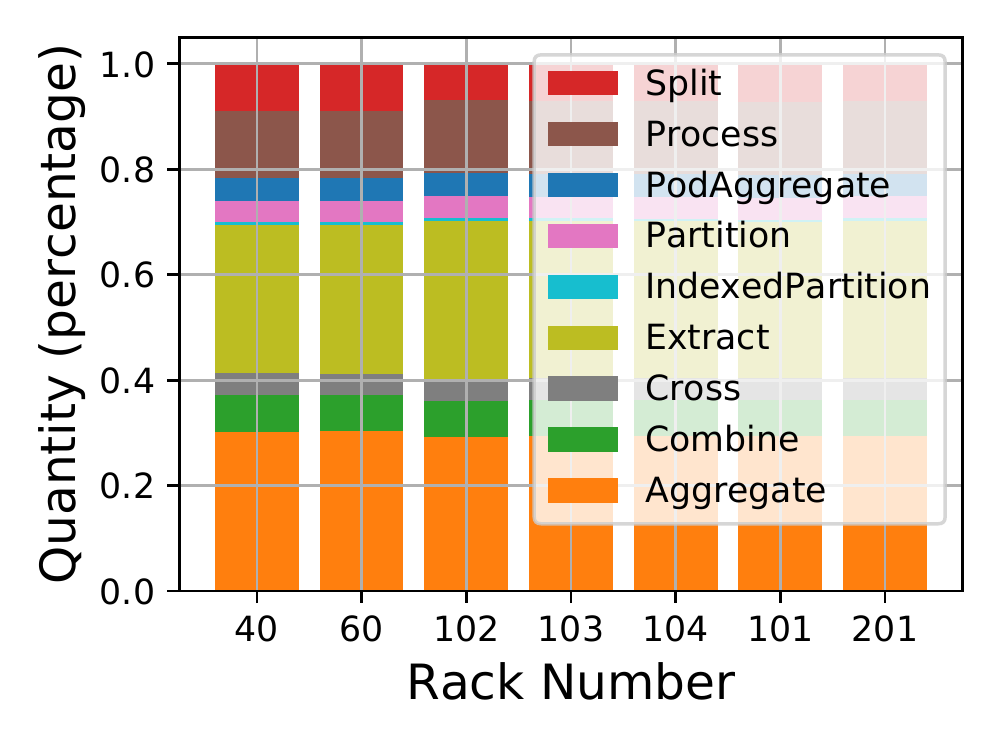}
	\end{minipage}%
	\begin{minipage}[b]{0.5\columnwidth}
		\includegraphics[width=\textwidth]{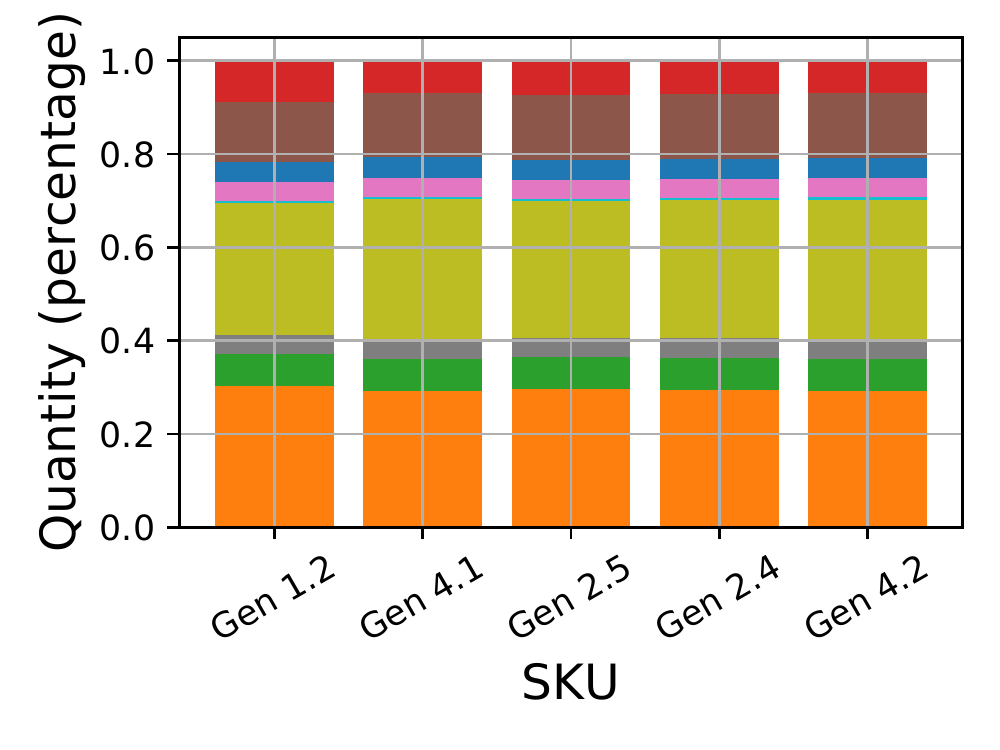}
	\end{minipage}%
	\vspace{-0.3cm}
	\caption{Task type distributions across racks (left) and SKUs (right) are very similar, indicating that, at the aggregate level, machines are fairly receiving a similar combination of workloads that is representative to the whole cluster.}\label{fig:type}
	\vspace{-0.2cm}
\end{figure}

\mypar{Machine-level abstraction (Level IV)} This abstraction is based on the observation made by the DS that the scheduler randomizes tasks uniformly across nodes, as shown in the left of \autoref{fig:type}. This allows us to focus directly on optimizing machine behavior, ignoring task-to-task interactions. Again this is held true in a statistical sense. Fortunately, big-data systems are by design very resilient to individual failures or misbehaving nodes, thus a statistical improvement is all we care for.

\mypar{Machine group-level abstraction (Level V)} The final observation from the DS is that tasks are also spread uniformly across SKUs (see the right of \autoref{fig:type}).
This allows us to model at the machine group-level, ignoring specific machine-level effects.

Level V is sufficiently abstract to allow us to model the whole infrastructure in a tractable yet comprehensive manner. In particular, we can now only focus on machine group-level metrics---examples of such metrics are shown in \autoref{tab:metrics}.
Such metrics can be directly affected by machine group-level parameters and the resulting impact can be mathematically quantified.

\begin{table}[h]
	\caption{Examples of performance metrics at the machine group level}\label{tab:metrics}
		\vspace{-0.2cm}
			\begin{tabular}{p{2.2cm}p{3.6cm}p{1.6cm}}
		\toprule
		Name      & Description & Affected System Metrics                              \\ \midrule
		Total Data Read      & Total bytes read per hour per machine       & Throughput rate\\
		Number of Tasks    & Total number of tasks finished per hour per machine      & Throughput rate        \\
		Bytes per Second    & Ratio of sum of the total data read and total execution time per machine      & Throughput rate        \\
		Bytes per CPU Time    & Ratio of sum of the total data read and total CPU time per machine      & CPU processing rate        \\
		CPU Utilization & Time-average CPU utilization per hour in percentage         & Utilization level         \\
		Average Running Containers & Time-average running containers per hour    & Utilization level         \\	
		\bottomrule
	\end{tabular}
		\vspace{-0.3cm}
\end{table}

\section{System Overview}\label{sec:overview}

In this section, we present \sysname's architecture (Section~\ref{ssec:archi}) and the three tuning approaches supported by \sysname (Section~\ref{ssec:tuningmodes}).

\begin{figure}[!t]
	\centering
	\includegraphics[width=\columnwidth]{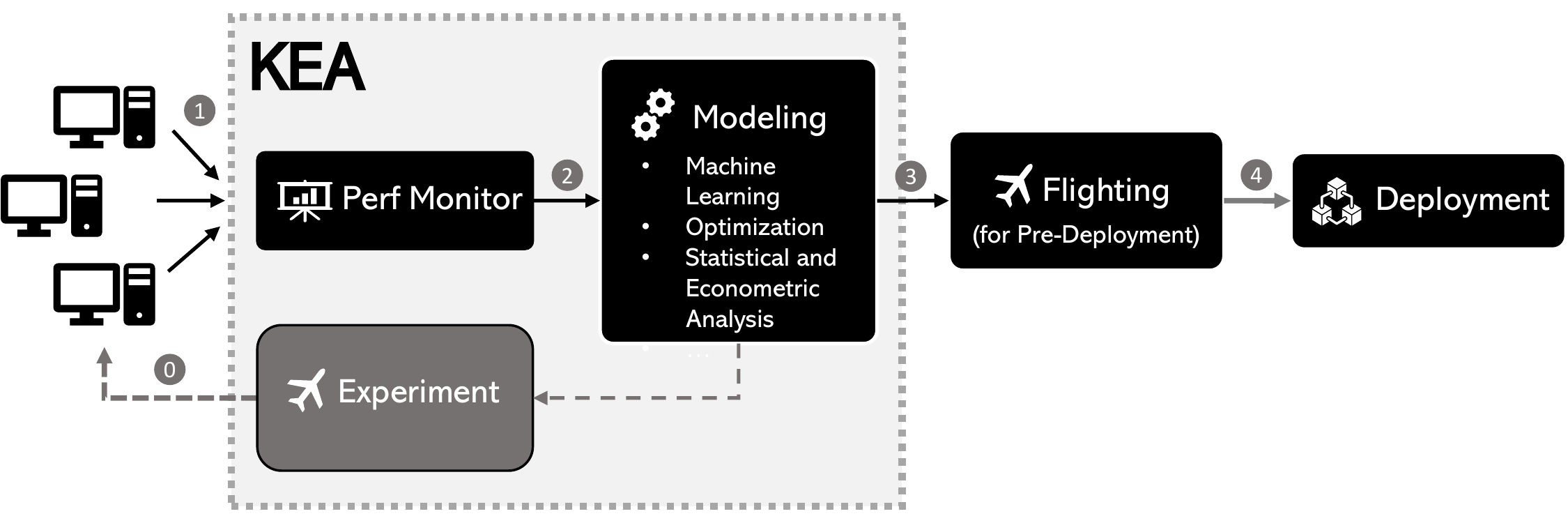}
	\vspace{-0.5cm}
	\caption{Overall architecture of KEA. The Experiment module and the flighting module share the same infrastructure to deploy new configurations. The Experiment module is only used for \ttwo.}\label{fig:overview}
	\vspace{-0.0cm}
\end{figure}

\subsection{Architecture}
\label{ssec:archi}

\sysname's architecture is shown in \autoref{fig:overview}. It consists of the following three main modules.

\begin{figure}[!t]
	\centering
	\includegraphics[width=0.96\columnwidth]{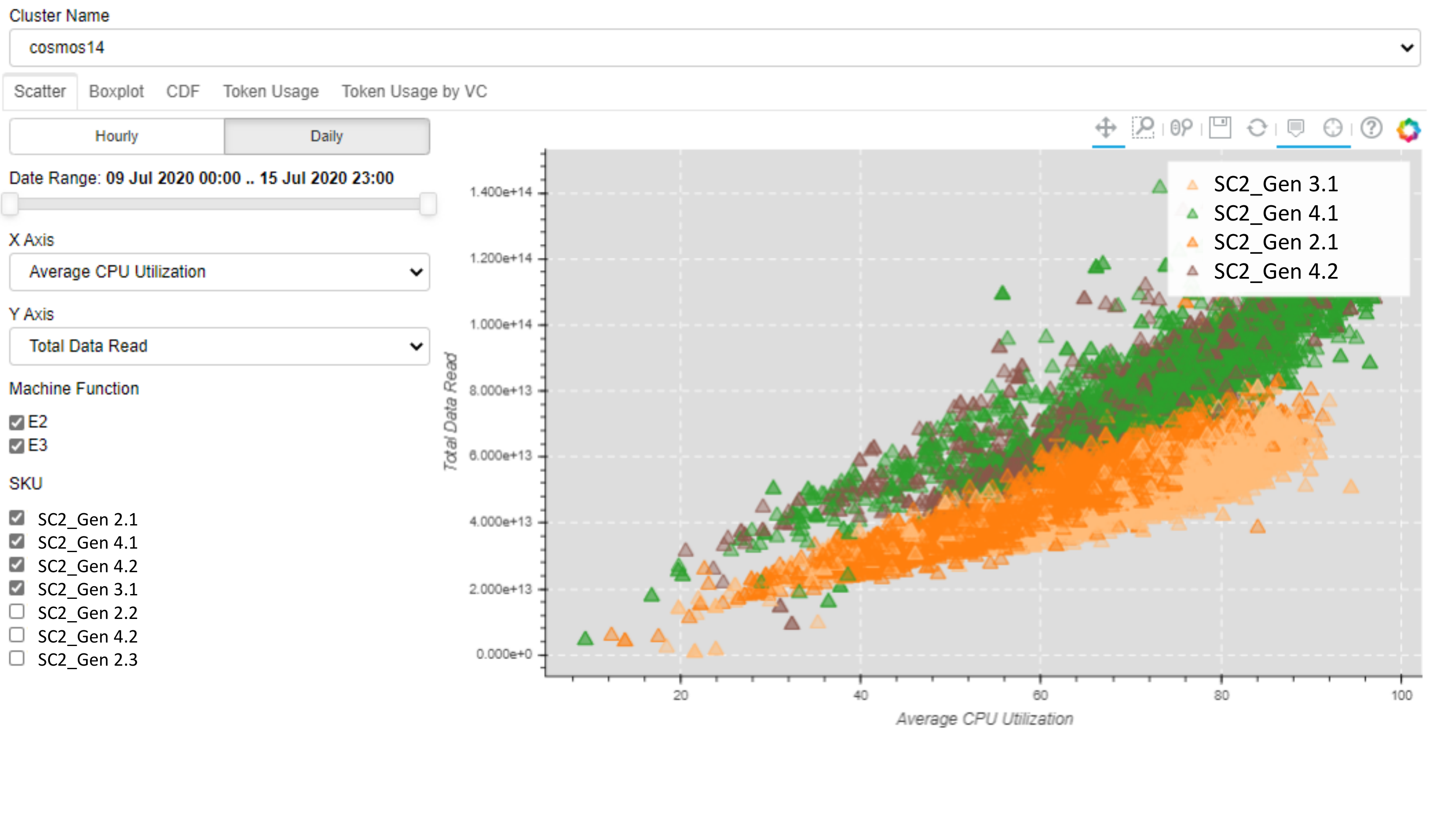}
	\vspace{-0.6cm}
	\caption{Scatter view of the performance monitor. We observe a linear trend between the total throughput (Total Data Read in the y-axis) and the  machine CPU utilization level (x-axis). The distribution varies across machine groups.}\label{fig:scatter2}
	\vspace{-0.4cm}
\end{figure}

The \textbf{Performance Monitor} joins data from various Cosmos sources and calculates the performance metrics of interest, providing a fundamental building block for all the analysis. 
To collect and prepare these metrics at a daily basis, an end-to-end data orchestration pipeline is developed and deployed in production on Cosmos itself. 
Figure~\ref{fig:scatter2} shows an instance of our visualization dashboard using the processed data. The \textit{scatter view} depicts the data in a disaggregated way with each point corresponding to one observation for a machine during one hour.

The \textbf{Modeling Module} proposes the optimal configurations as described in Phase II of Figure~\ref{fig:rcp} (see \autoref{sec:rcp}). Depending on the application, different methods can be used, such as ML, optimization, statistical analysis, or econometric models.

The \textbf{Experiment Module} is used to perform experiments on a group of machines to gather performance data for new configurations, when this is required to model the system.

The \textbf{Flighting Tool} facilitates the deployment of configuration changes to any machine in the production cluster as a safety check before performing the full cluster deployment using the \textbf{Deployment Module}. In the flighting tool, users can specify the machine names and the starting/ending time of each flighting and create new builds to deploy to the selected machines.

As the modeling and experiment modules vary across different applications, they are discussed in more detail for each application in Sections~\ref{sec:tuning}--\ref{sec:pc}.

\subsection{Tuning Approaches}
\label{ssec:tuningmodes}

We now describe the different tuning approaches enabled by \sysname.

Our goal is to avoid, to the degree possible, ``experimental tuning'', i.e., the modeling of the system based on rounds of experiments that are costly and often prohibitive, given the scale and constraints of our infrastructure. 
To this end, we make the crucial observation that \emph{in many scenarios we can model the effect of new configurations to the system's behavior based solely on data collected from existing cluster operating points}. For example, to determine the maximum number of tasks to run per machine, we can simply look at the performance of the system when different numbers of tasks happened to run on those machines (given that the number of tasks per machine changes over time based on the cluster load).
Note that in cases of future cluster planning (e.g., how much memory to use for future machines), experimental tuning is not even an option---relying on existing cluster operation points is the only feasible way.
However, some applications require additional exploration to actually deploy the new configuration and evaluate the observed performance---those usually involve configurations that the system has not seen before. To this end, we introduce the following three tuning approaches:

\myparr{\tone} aims at improving the operation efficiency of an existing fleet of machines. It can be applied to configuration scenarios for which the telemetry data collected from past cluster operation is sufficient to predict the performance of a new configuration, thereby completely bypassing experiments. 
Based on the modeling results, we only need to conduct flighting as a pre-deployment safety check to validate the model prediction.
The \sysname modules (see \autoref{fig:overview}) required for this type of tuning are: (1) performance monitor, (2) modeling, (3) flighting, and (4)~deployment.

\myparr{\tthree} aims to support planning for future cluster machines. It also uses modeling based on existing cluster operational points. 
In this case, there is no flighting nor actual deployment, since the output is configuration of machines that do not yet exist.
Applications fueled by this tuning approach focus on forecasting for future scenarios, such as choosing memory and CPU for future machines.
The required \sysname modules here are: (1) performance monitor and (2) modeling.

\myparr{\ttwo} is used when the current telemetry data is not sufficient for predicting the performance of a new configuration. Used as our last resort (due to its increased cost), it strategically performs experimental deployments of new configurations to gather new cluster operational points. The data collected during experiments are then used to model the system. This approach uses all \sysname modules in \autoref{fig:overview}.

\begin{table}[t]
	\caption{\sysname applications}\label{tab:app}
	\begin{tabular}{p{2.1cm}p{1.9cm}p{3.2cm}}
		\toprule
		Application      & Tuning \linebreak Approach & Tuning Parameter                               \\ \midrule
		YARN \linebreak Configuration      & \tone       & Maximum running containers for each SC-SKU combination\\
		SKU Design       &\tthree      & Amount of RAM, SSD per machine             \\
		Power Capping    & \ttwo      & \% below current provision level        \\
		Software \linebreak Configurations & \ttwo         & Binary choice between \verb|SC1| and \verb|SC2|         \\
		\bottomrule
	\end{tabular}
	\vspaceafigure
\end{table}

\vspace{1mm}

By employing these tuning approaches, \sysname can tackle a wide variety of configuration scenarios that we encounter in Cosmos. In this paper, we focus on four scenarios that we have already addressed using \sysname in production. Optimizing these scenarios, which are summarized in \autoref{tab:app} and described below, has led to over tens of millions of dollars yearly savings for the company.

\mypar{(1) YARN Configuration Tuning} The goal is to re-balance the workload across different groups of machines and achieve better cluster throughput by tuning YARN configuration parameters. Specifically, we focus on the \verb|max_num_running_containers| parameter that can be specified for each machine group corresponding to a software-hardware combination. This parameter limits the maximum number of containers that can be executed simultaneously on a machine. \verb|max_*|/\verb|min_*| is a commonly seen parameter in various configuration settings, and \sysname can be employed to tune many such parameters.

\mypar{(2) Machine Configuration Design} 
The goal is to determine the hardware configuration (amount of RAM and SSD size) of future machines in a way that does not leave any of the resources idle and without impacting throughput either.
The most cost-efficient configuration tailored to current customer workloads is desired.

\mypar{(3) Power Capping} The bottleneck for fitting more machines in a rack or in a data center is provisioned power, not space.
With power capping we focus on determining new power limits. Originally, the Cosmos machines were provisioned with a conservatively high power consumption limit---based on years of observation, this power limit turns out not to be cost-effective. 
By capping the power utilization and provisioning less power per machine, we are able to increase the number of machines per rack and thus per data center. This way the fixed costs of racks and data centers, which account for a large portion of a data center cost~\cite{thedatacenter}, are amortized across more machines.

\mypar{(4) Selecting Software Configurations (SCs)} Machines are purchased with different hardware (i.e. SKU) and deployed with different software versions. In Cosmos we use two main software configurations, denoted here as \verb|SC1| and \verb|SC2|, corresponding to different mappings of logical drives to physical media. We use \sysname to determine which of the two options is preferable.

Application (1) can be tackled based solely on existing observational data that allow us to build models and predict performance for different maximum number of running containers per SC-SKU combination. Hence, it is a good example of \tone. Application~(2) belongs to \tthree, as it focuses on future planning of the system. Finally, for applications~(3) and~(4) we use \ttwo, because without actual experiments with new configurations, it is very difficult to predict the impact of new power capping limits or new software configurations.

\section{\tone}\label{sec:tuning}

In this section, we develop an \tone approach to tune the configurations by building the predictive models (see Phase II in Figure~\ref{fig:rcp}) to avoid the need of rounds of cluster-wide experiments, which are required by the black-box approaches, such as Bayesian optimization (BO) and reinforcement learning (RL)~\cite{genkin2016automatic,wager2018estimation,zhang2019end}. We find that cluster-wide experiments are impractical in large-scale production environments like Cosmos because of how slowly and carefully changes to production need to be made. By properly modeling the dynamics of the system, we can predict the potential performance changes with different configurations.

\subsection{Modeling}\label{sec:ttwo} 
The \tone approach consists of two modules: (1) the \textit{What-if Engine} to predict the performance metrics given different configurations and (2) the \textit{Optimizer} to select the optimal solution. 

In the What-if Engine, we aim to predict the resulting performance given a new set of configurations. On the path to tackle this problem, we make two crucial observations:
\begin{itemize}
	\item \textit{The change of a particular set of configuration parameters usually affects one (or a few) sets of metrics directly, and the impact is easy to measure (see Phase II in Figure~\ref{fig:rcp}).} For instance, by changing the configuration for the maximum number of running containers, the metrics directly impacted are the actual running containers of a machine and its distribution. It is clear that reducing the max will shift its distribution towards the lower end. 
	\item \textit{We can capture the dynamics between different sets of parameters and better understand how the change in one set of metrics affects the others using ML models, i.e. the chain-effect.} In the observational data, due to the natural variance of the system operation, we have a full-spectrum of the ranges of the performance metrics (see Figure~\ref{fig:scatter2} where we have observations for machines running with various levels of CPU utilization). Based on this variation, we can develop models to mimic the dynamics between different sets of metrics and map one metric to another.
\end{itemize}

The above two observations are important building blocks to capture the relationship between changes in configurations to the changes in the objective functions (or constraints) that we hope to optimize upon. In summary, we use the following three step process: (1) based on the set of parameters that we hope to tune, we identify the set(s) of metrics that will be directly impacted; (2) we build ML models to understand how this set of metrics affects the others, especially the ones that relate to our objective functions/constraints; and (3) based on the resulting formulation, we perform optimization to pick the optimal configuration (see Phase II in Figure~\ref{fig:rcp}).

For the development of the ML models, we conclude that the dynamics between the different sets of metrics remain the same, even with different configuration settings. Those system aspects reflect the mechanics of the infrastructure and the characteristics of the workloads.
For instance, in Figure~\ref{fig:scatter2}, even with different levels of CPU utilization or the workload levels, the \textit{relationship} between the resulting throughput and the CPU utilization level can be expressed with the \textit{same} formulation for each group of machines with a particular software-hardware combination. This relationship will not be affected by the external configuration, such as YARN configuration settings. In this research, we leverage those system fundamentals to predict the resulting performance under new configurations to avoid the need for experiments. Those are the calibrated models we want to develop in Phase II as in Figure~\ref{fig:rcp}.

Based on the observational data, sets of ML models can be built, such as $g_k(\cdot)$, $h_k(\cdot)$ and $f_k(\cdot)$, for each SC (software configuration)-SKU (hardware) combination $k$, to capture the relationship between the different sets of metrics, such as (1) the number of running containers versus the CPU utilization level, (2) the CPU utilization level versus the number of tasks finished per hour, and (3) the CPU utilization level versus the task latency respectively:
\begin{align} 
x_k = g_k(m_k)~~~\forall k=1,2,3,\cdots,K  ,\label{eq:test11}\\ 
x'_k = g_k(m'_k)~~~\forall k=1,2,3,\cdots,K  ,\label{eq:test22}\\ 
l_k = h_k(x_k)~~~\forall k=1,2,3,\cdots,K  ,\label{eq:test33}\\ 
l'_k = h_k(x'_k)~~~\forall k=1,2,3,\cdots,K  ,\label{eq:test44}\\ 
w_k = f_k(x_k)~~~\forall k=1,2,3,\cdots,K  ,\label{eq:test55}\\ 
w'_k = f_k(x'_k)~~~\forall k=1,2,3,\cdots,K  ,\label{eq:test66}
\end{align}
where,

\begin{tabular}[H]{lp{6cm}}
	$k$:& the index for the SC-SKU combination, $k=1,2,3,\cdots,K$.\\
	$m_k$:& the number of running containers (simultaneously) per machine with SC-SKU combination $k$. 	\\
	$m'_k$:& the original average number of running containers per machine with SC-SKU combination $k$. 	\\
	$x_k$ and $x'_k$:& the CPU utilization level for machines with SC-SKU combination $k$, given the number of running containers $m_k$ and $m'_k$ respectively. \\
\end{tabular}

\begin{tabular}[H]{lp{6cm}}
	$l_k$ and $l'_k$:& the number of tasks finished per hour on a machine with SC-SKU combination $k$, given the CPU utilization level $x_k$ and $x'_k$ respectively.\\
	$w_k$ and $w'_k$:& the average task latency for machines with SC-SKU combination $k$, given CPU utilization level $x_k$ and $x'_k$ respectively.\\
\end{tabular}

Many other ML models can be developed to involve a larger set of metrics of interest, such as the resource utilization of SSD, RAM (used in Section~\ref{sec:skuDesign}), or network bandwidth, for different applications. 
In general, we use regression models as the predictors, such as linear regression (LR), support vector machines (SVM), or deep neural nets (DNN). Linear models are more explainable, which is critical for domain experts.

For a complex system, given the fact that there are only a handful number of machine groups with different software-hardware combinations, based on the needs of different projects, a small number of models per group are sufficient to mimic the full dynamics of the system, which is tractable and easy to maintain.

In the Optimizer, different objective functions and constraints can be easily plugged-in with respect to the goals of the applications, and the corresponding ML models with respect to the directly impacted performance metrics and the ones related to the objective functions/constraints can be used.
 
\subsection{Yarn Configuration Tuning}
For the application of tuning the maximum running containers in YARN, after Phase I, we decide to maximize the total number of running containers (relates to the sell-able capacity for the cluster) subject to the same overall average task latency at the cluster level as in the current operation. Therefore, the directly impacted performance metric is the number of running containers on the machine. And we maintain the same level of task latency (cluster-wide average) as the constraint. Thus, the optimization problem can be formulated with a closed-form objective function as a linear programming (LP) problem:
\begin{align} 
\max_{m_k, k =1,2,3,\cdots,K} ~ &\sum_{k =1,2,3,\cdots,K}^{}{m_kn_k},\label{eq:obj}  \\
\textrm{s.t.} ~ \bar{W} &\leq \bar{W'},\\ 
\bar{W} &= \frac{\sum_{k}^{}{w_kl_kn_k}}{\sum_{k}^{}{l_kn_k}},\\
\bar{W'} &= \frac{\sum_{k}^{}{w'_kl'_kn_k}}{\sum_{k}^{}{l'_kn_k}},\\
\eqref{eq:test11}&-\eqref{eq:test66}.\nonumber
\end{align}
where,

\begin{tabular}[H]{lp{6cm}}
$n_k$:& the number of machines in the cluster for machine function-SKU combination $k$.\\
	$\bar{W}$ and $\bar{W'}$:& the overall average task latency for the full cluster, given CPU utilization level $x_k$ and $x'_k$ respectively, calculated as the weighted average of task latency running on different groups of machines.\\
\end{tabular}

The optimal solution of the optimization, $m_k^*~\forall k=1,2,3,\cdots, K$, can be obtained by commercial solvers, indicating the optimal workload distribution across different groups of machines. Based on the changes of the workload distribution, we can modify the configuration for the maximum running containers accordingly, increasing or decreasing it for different SC-SKU combinations. In the production system, we are also conservative about the changes in configuration. Therefore, extra constraints can be added to limit the range of the decision variables.

In the current system, we observe that the tasks running on slower machines are more likely to be the stragglers that slow down a job. 
The re-balancing of the workloads suggested by the model reduces the workload skew and shifts traffic from slower machines to faster machines to improve the overall efficiency. With the increased utilization level on faster machines, we expect mild performance degradation for their tasks. However, those are less likely to be on the critical path of a job that directly impacts the job-level latency (see Figure~\ref{fig:critical}). Therefore, even though the constraint for the optimization formulation ensures the same average task-level latency, the tuning improves the performance of the straggler tasks thus improving the job-level latency.

\subsubsection{Modeling Results}\label{sec:res}

We present the training and tuning results from one of our production clusters with over 45,000 machines for a seven day period. The full KEA pipeline was set up to collect data on a daily basis and propose a new configuration each day.

\begin{figure}[!t]
	\centering
	\begin{minipage}[b]{0.5\columnwidth}
		\includegraphics[width=\textwidth]{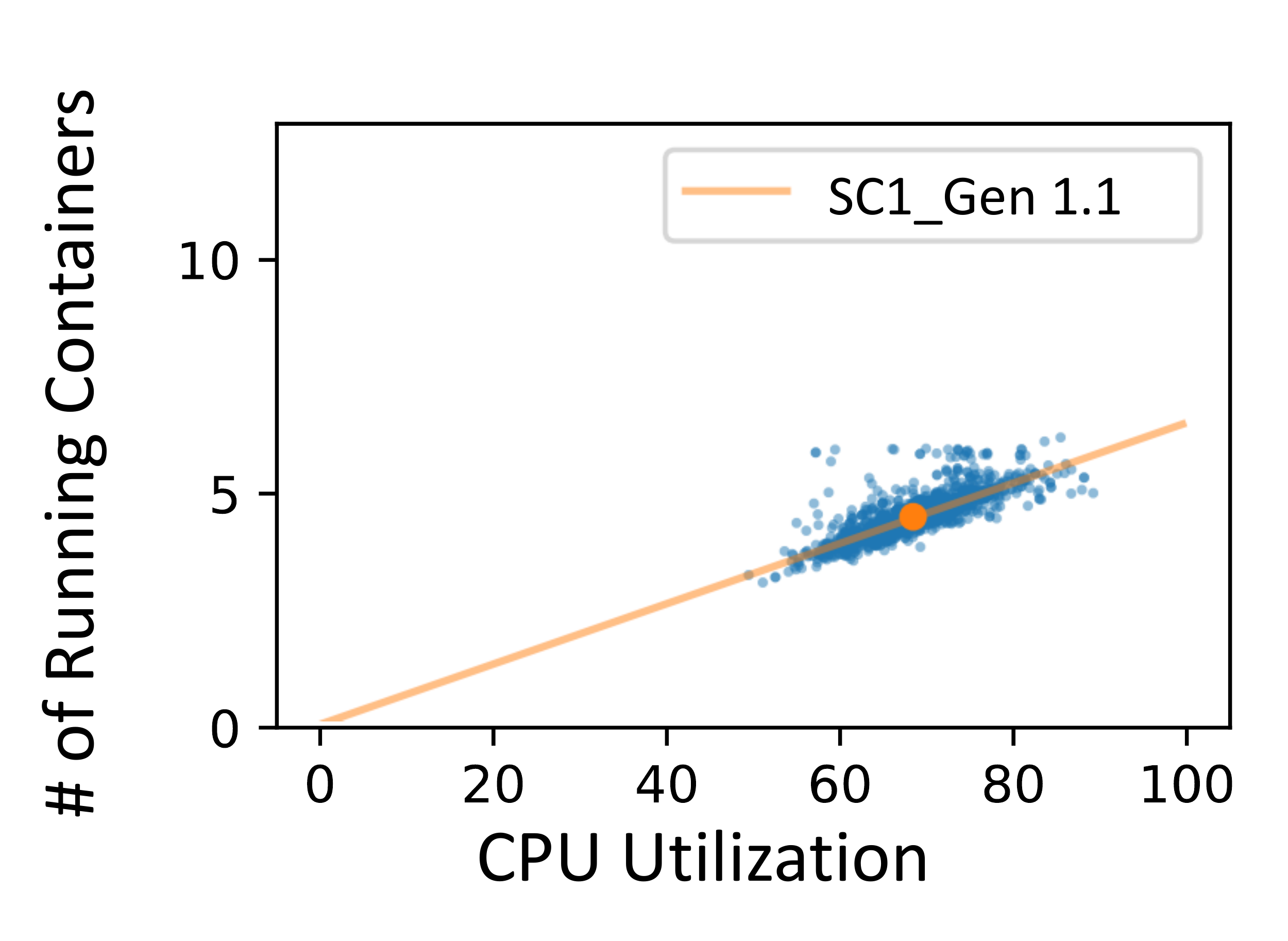}
	\end{minipage}%
	\begin{minipage}[b]{0.5\columnwidth}
		\includegraphics[width=\textwidth]{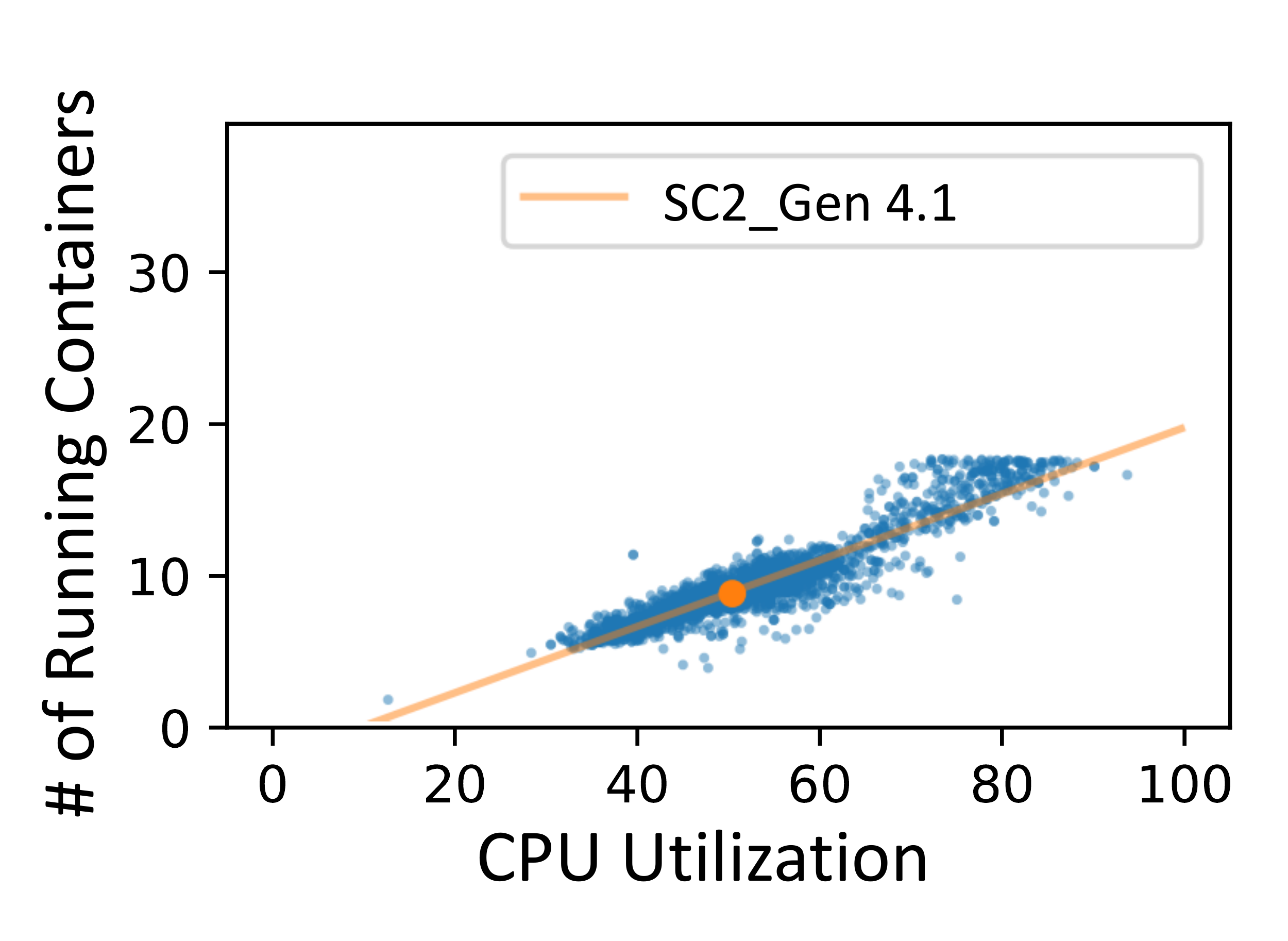}
	\end{minipage}%

	\begin{minipage}[b]{0.5\columnwidth}
		\includegraphics[width=\textwidth]{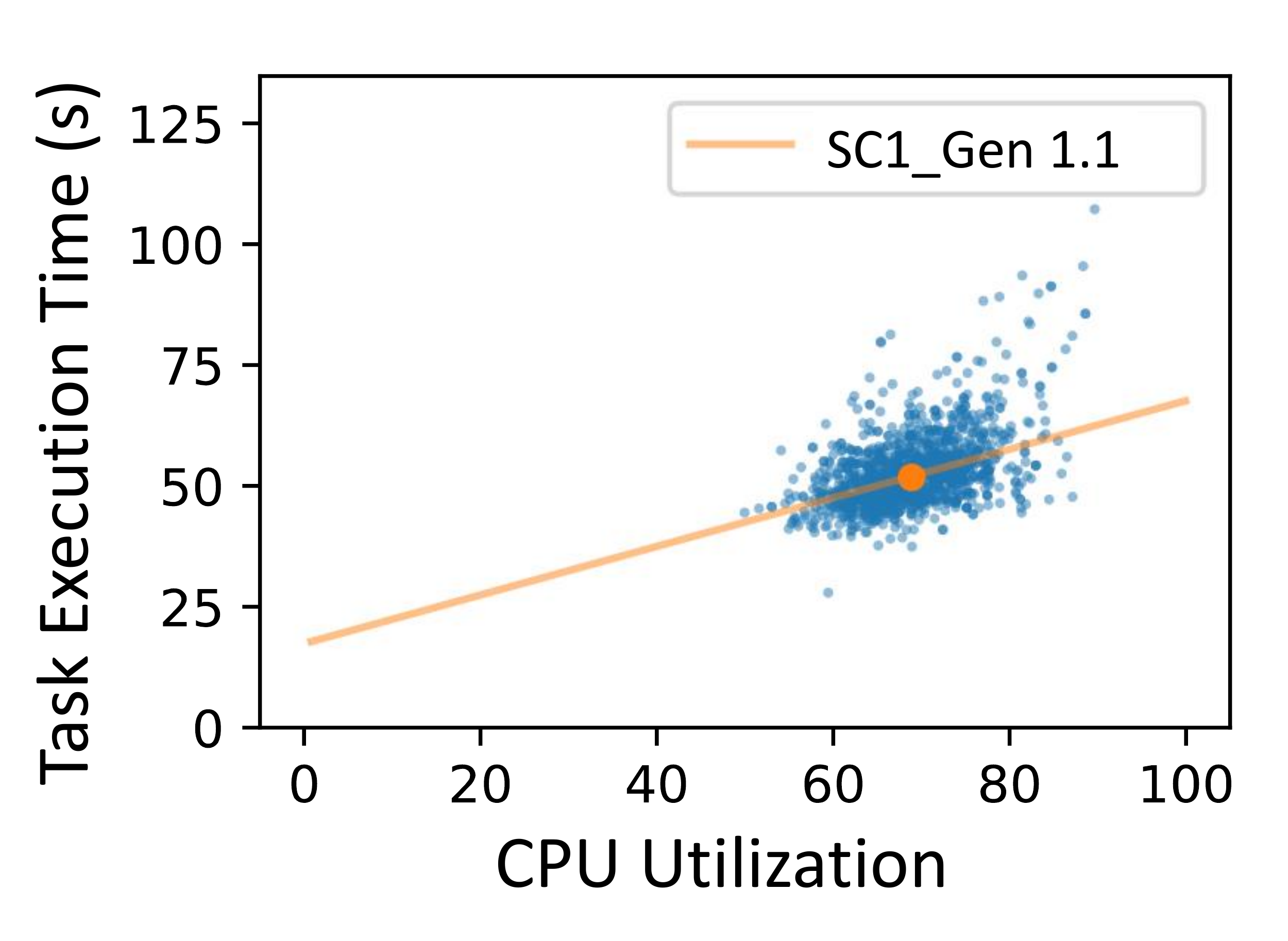}
	\end{minipage}%
	\begin{minipage}[b]{0.5\columnwidth}
		\includegraphics[width=\textwidth]{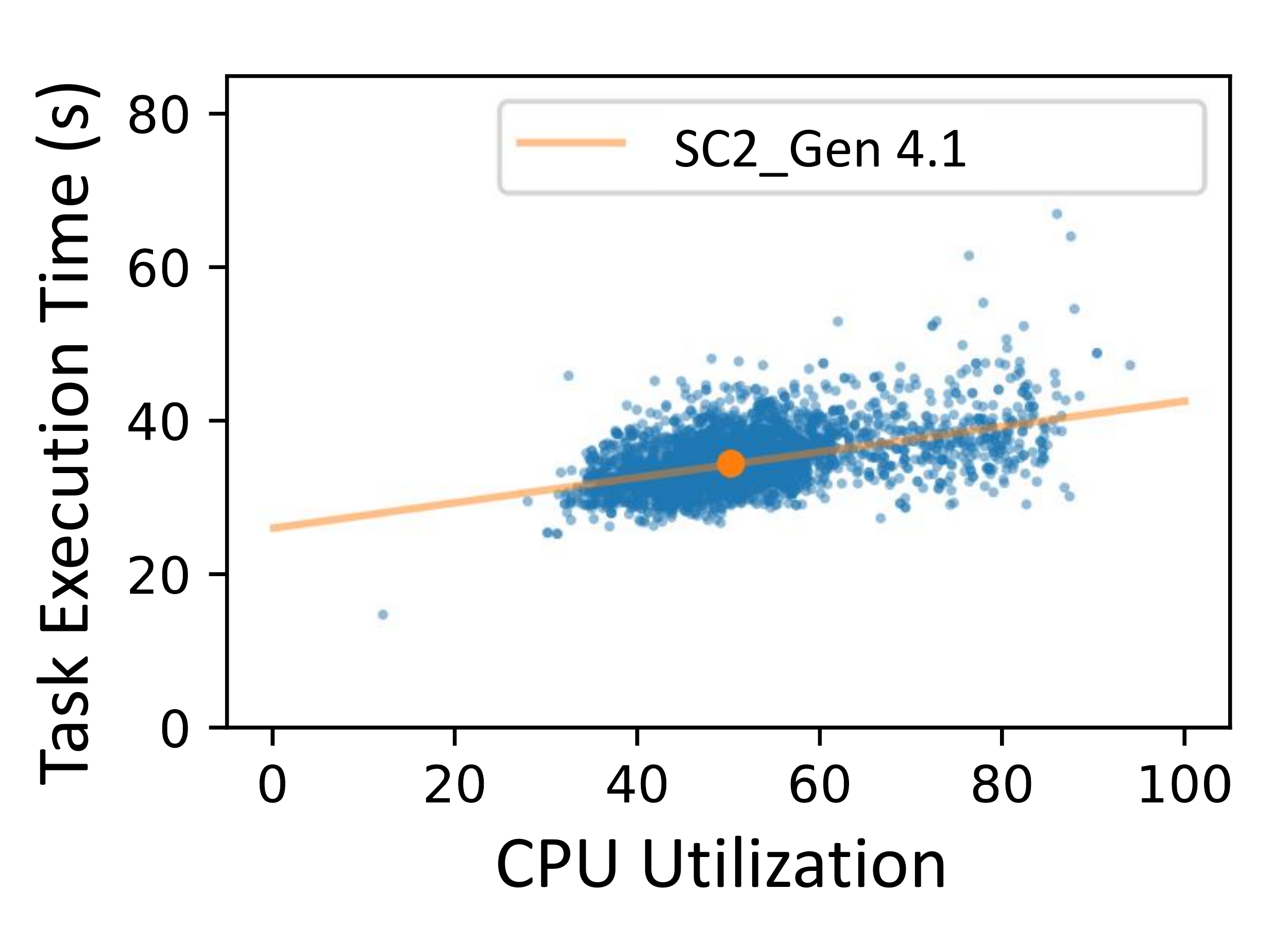}
	\end{minipage}%
	\vspace{-0.3cm}
	\caption{The set of calibrated models for different SCs and SKUs}\label{fig:vts}\label{fig:exe}\label{fig:ctn}
	\vspace{-0.4cm}
\end{figure}

We used a Huber Regressor for the prediction of the set of performance metrics in the What-if Engine, which is more robust to outliers compared to the
Least Squares Regression~\cite{owen2007robust}.
Figure~\ref{fig:ctn}
shows the set of calibrated models (see Phase II in Figure~\ref{fig:rcp}) to depict the running containers and task execution time in seconds versus CPU utilization level. Each small dot corresponds to an observation aggregated at the daily level for a machine. The line shows the model estimation. The large dot in the center of the figure indicates the median level of the variables across all observations. 

\begin{figure}[!t]
	\centering
	\includegraphics[width=0.36\textwidth]{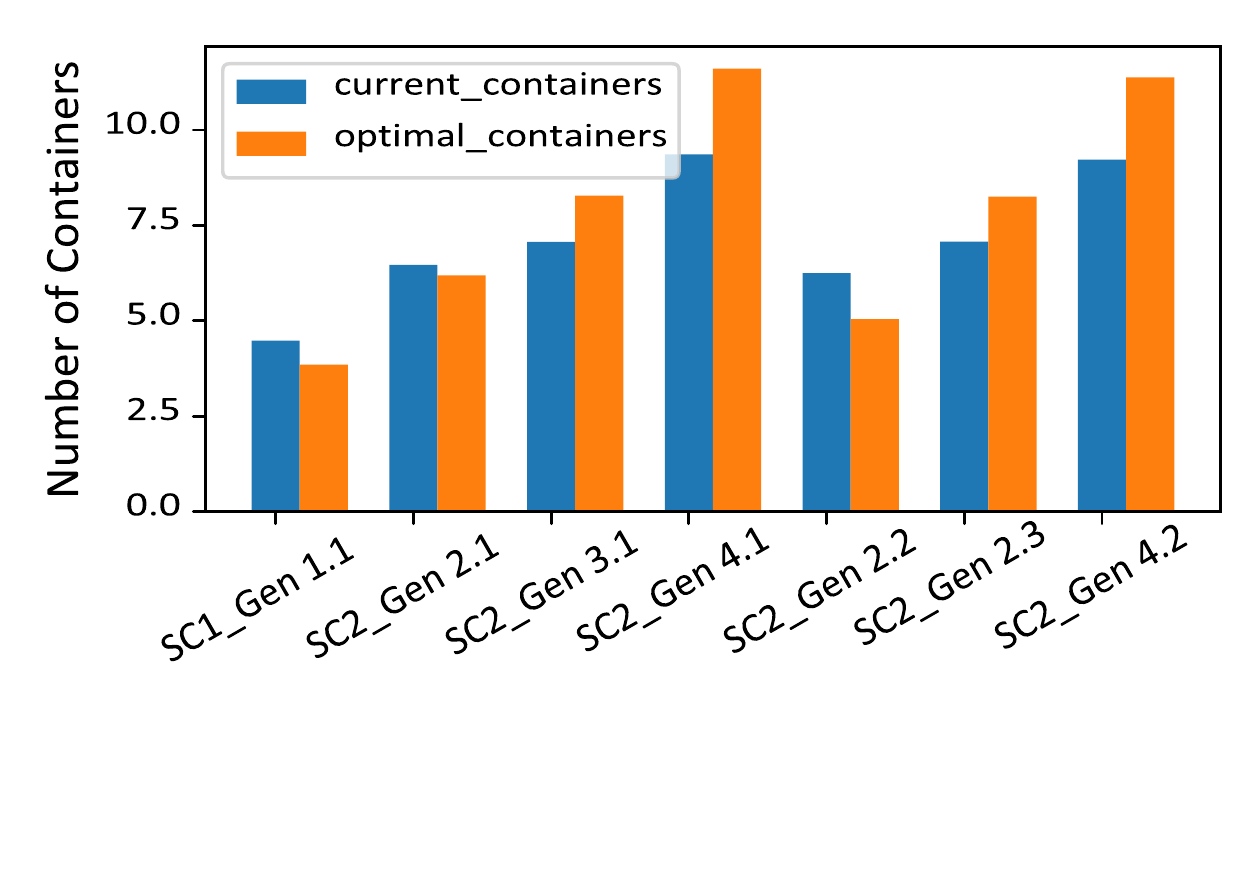}
	\vspace{-1.3cm}
	\caption{Suggested configuration change}\label{fig:opt1}
	\vspace{-0.4cm}
\end{figure}

Figure~\ref{fig:opt1} shows the optimization results in terms of the suggested shift of current workloads (calculated as the number of containers running per machine). For slower machines, such as \verb|Gen 1.1|, the model suggests to decrease the utilization by reducing the number of running containers, while for faster machines, such as \verb|Gen 4.1|, the model suggests to increase it. 
We also ran the same optimization model focusing on a higher percentile of CPU utilization level, corresponding to the situation where the whole cluster is running with heavy workloads. The suggested configuration change is the same in terms of the direction for the gradients. We conclude that for the cluster when running with low/median/heavy workload, the same configuration change is desired.

\subsubsection{Flighting and Deployment}\label{sec:eval}
The flighting module is one of the most important components for KEA that leads to its applicability to large production systems. Before our first deployment, to establish confidence and demonstrate robustness, we did five rounds of flighting where we validated the possibility of increasing maximum running containers for different SKUs to increase utilization. 

The first pilot flighting was on 40 \verb|Gen 1.1| machines to confirm that reducing the \verb|max_num_running_containers| in the YARN configuration file is affecting the real observed maximum number of running containers. The second piloting flighting experiment was on \verb|Gen 4.1| machines to confirm that increasing the \verb|max_num_running_containers| in the YARN configuration is effective and allows the machines to run more workloads. The third piloting experiment was on two sub-clusters~\cite{curino2019hydra} of machines (each with around 1700 machines) to validate if the updated configuration changes the workload distribution. The fourth pilot flighting was for three sub-clusters of machines and validated the benefits of tuning, i.e. adding more containers to the sub-cluster with better performance.

The production roll-out process is very conservative where we only modify the configuration by a small margin, i.e. decrease or increase the maximum running containers for each group of machines by one. We extracted the performance data for the periods of one month before and one month after the roll-out. 
We use \textit{treatment effects} to evaluate the performance changes~\cite{holland1986statistics} during the two periods with significant tests. 
We observe that with the same level of latency (measured by average task latency), the throughput (measured by Total Data Read) is improved by ~9\%. 

\begin{figure}[!t]
	\centering
	\includegraphics[width=\columnwidth]{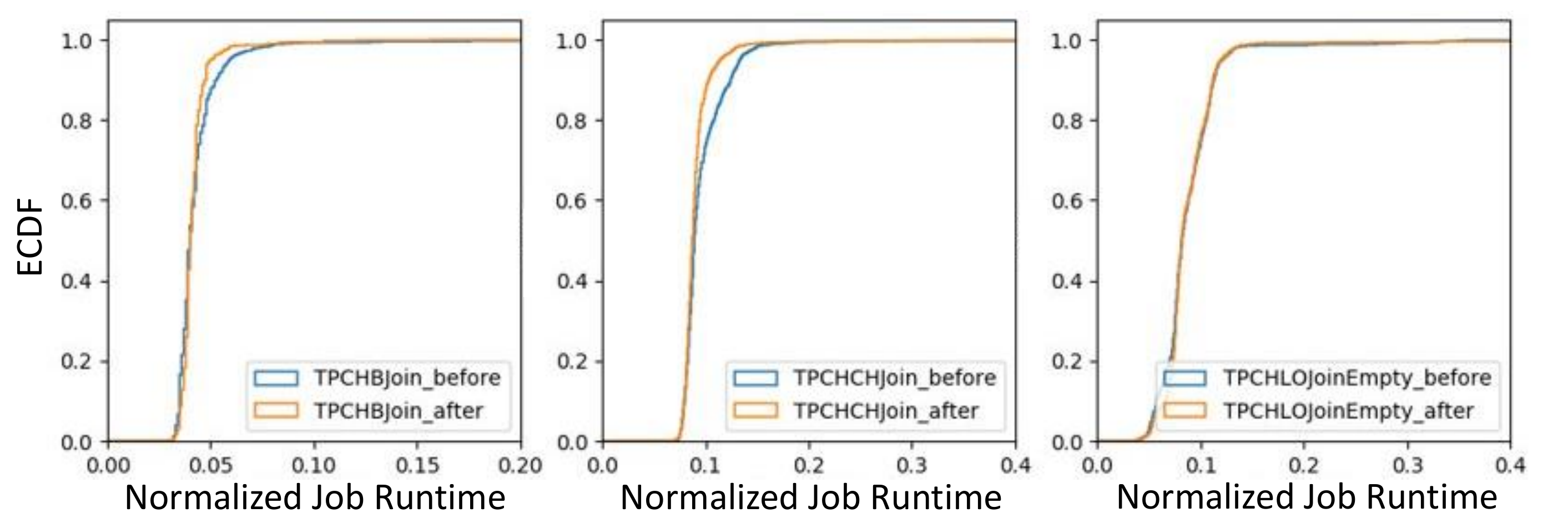}
	\vspace{-0.6cm}
	\caption{{\color{black}Job runtime distribution for 3 benchmark jobs before and after KEA deployment}}\label{fig:RJobs}
	\vspace{-0.4cm}
\end{figure}

{\color{black}We also evaluate a set of benchmark jobs originating from TPC-H and TPC-DS before and after KEA deployment. The average job runtime in this case was improved by 6\% (see Figure~\ref{fig:RJobs}).}

For this round of deployment, conservatively, we gain ~2\% sell-able capacity from the cluster (measured by the total number of containers with the same level of latency as before) by only modifying the maximum running containers for each SKU-SC by one. 
This is equivalent to tens of million dollars cost saving per year. We are expecting 5\% more over the existing capacity in the next round where we are allowed to modify the maximum running containers by two. 
The t-value for the Student's t-test~\cite{student1908probable} is 4.45 and 7.13 respectively, indicating that the resulting difference in the data distributions for the before and after periods is significant. 

\subsection{Discussion}
The What-if Engine can be extended to other performance metrics of interest by
identifying the most relevant sets of performance metrics and modeling the dynamics between them using predictive models. 

\begin{figure}[t]
	\centering
	\begin{minipage}[b]{0.5\columnwidth}
		\includegraphics[width=\textwidth]{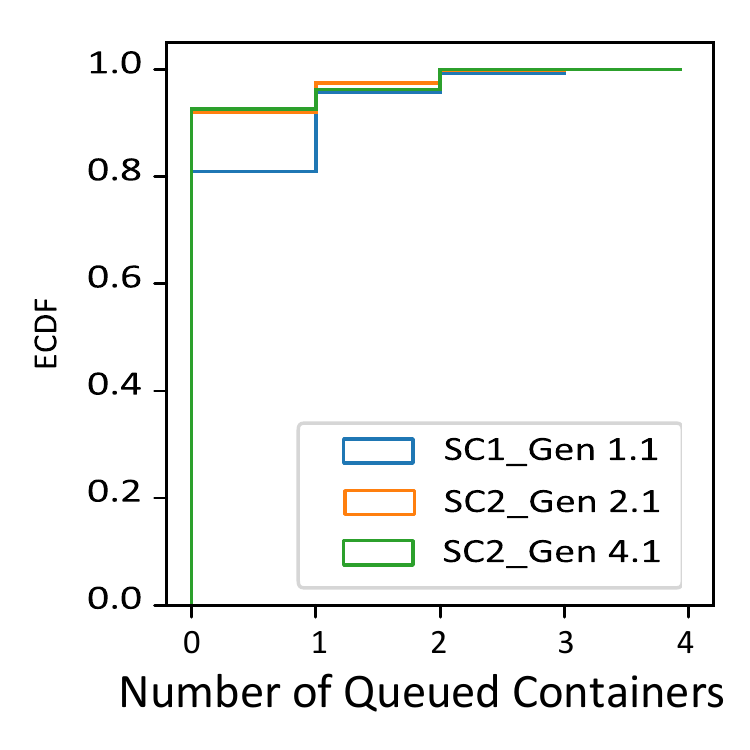}
	\end{minipage}%
	\begin{minipage}[b]{0.5\columnwidth}
		\includegraphics[width=\textwidth]{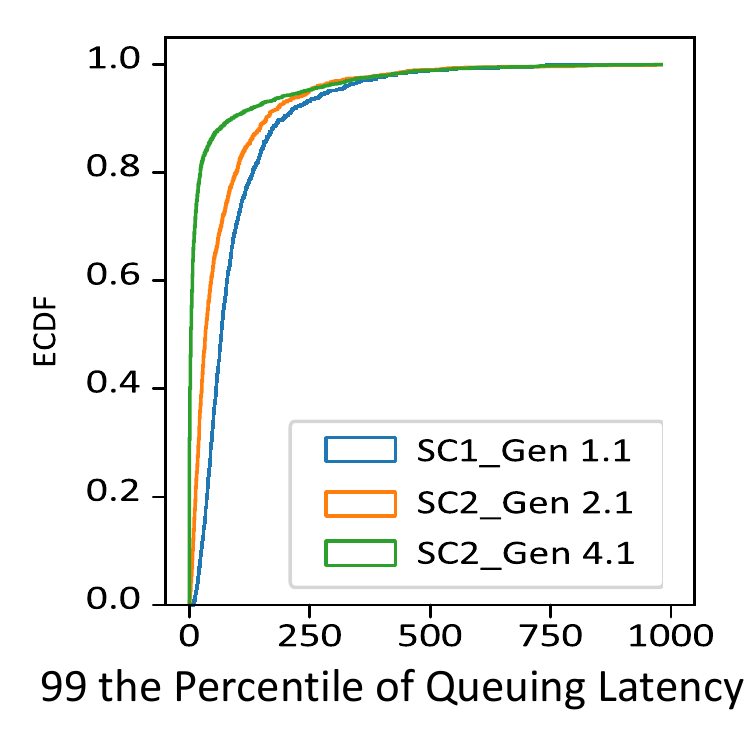}
	\end{minipage}%
	\vspace{-0.4cm}
	\caption{Number of queued containers (left) and 99$^{\textrm{th}}$ percentile of queuing latency in ms (right) for 3 SKUs}\label{fig:queuing}
	\vspace{-0.2cm}
\end{figure}

In the analyzed system, low priority containers will be queued on each machine when all machines in the cluster reach the maximum number of running containers. We observe that the queuing length and latency vary significantly for machines with different SKUs and SCs (see Figure~\ref{fig:queuing}). 
As faster machines have faster de-queue rate, we can allow more containers to be queued on them. Therefore, similar tuning methodology can be used to learn the relationship between the tuned parameters, i.e. the maximum queuing length, and the objective performance metrics, such as variance of queuing latency, to achieve better queuing distribution. 

Based on the ML models proposed in Equations \eqref{eq:test11}-\eqref{eq:test66}, KEA can also be used to convert any performance improvement into capacity gain (given the same task latency), allowing detailed quantitative evaluation for all engineering changes in monetary values.

\section{\tthree}\label{sec:skuDesign}

In this section, we discuss the applications focusing on scenarios for future planning. 
The same methodology for modeling is applicable as in Section~\ref{sec:ttwo}. The main focus is still to develop the predictive models in the What-If Engine to capture the relationship between the resulting objective functions with respect to a particular configuration change and let the Optimizer pick the optimal solution without deployments or experiments.

\subsection{Machine Configuration Design}

In this application, we focus on the resource utilization metrics of the machines (as opposed to the throughput or latency) that will influence decisions around what hardware components to purchase in future machines.
We have determined the CPU configuration for the next generation. In this work, we build a model to predict how much SSD and RAM we would need to buy given the number of CPU cores.

The different configurations for SSD and RAM directly impact their actual usage and distributions. Specifically, using ML models, we want to project the SSD and RAM usage as a function of the number of CPU cores used based on the observational data. Given a larger number of CPU cores available in the new generation, we can predict the usage of SSD and RAM based on the same functions:

\begin{align} 
s = p(c) = \alpha_s + \beta_sc,\label{eq:test77}\\ 
r = q(c) = \alpha_r + \beta_rc,\label{eq:test88}
\end{align}
where,

\begin{tabular}[H]{lp{6cm}}
	$c$:& number of CPU cores used in the observational data.\\
	$s$:& amount of SSD used when using $c$ CPU cores. 	\\
	$\alpha_s,\beta_s$:& parameters to predict the SSD usage. 	\\
	$r$:& amount of RAM used when using $c$ CPU cores. \\
	$\alpha_r,\beta_r$:& parameters to predict the RAM usage. 	\\	
\end{tabular}

In Equations~\eqref{eq:test77} and~\eqref{eq:test88}, for $p(c)$ and $q(c)$, we use a simple linear regression model. Based on the current data, we calibrate the values for $\alpha_r, ~\alpha_s,~\beta_r$ and $\beta_s$.

\begin{figure}[!t]
	\centering
	\begin{minipage}[b]{0.5\columnwidth}
		\includegraphics[width=\textwidth]{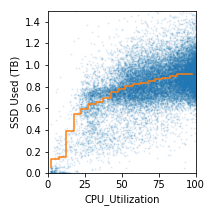}
	\end{minipage}%
	\begin{minipage}[b]{0.5\columnwidth}
		\includegraphics[width=\textwidth]{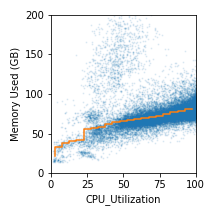}
	\end{minipage}%
	\caption{Resource utilization for SSD and RAM}\label{fig:sku}
	\vspace{-0.2cm}
\end{figure}

Figure~\ref{fig:sku} shows the current resource utilization for SSD and RAM with respect to different levels of CPU utilization for a particular SKU running with the production workloads. 
The observation is for each second for a full day with around 10.4 million records.

The $\alpha_s$ and $\alpha_r$ are the intercepts of the projection, indicating the SSD and RAM usage levels when running with $0$ cores. The $\beta_s$ and $\beta_r$ indicate the SSD usage per core and RAM usage per core.
A full distribution with regard to the $\alpha_s$, $\alpha_r$, $\beta_s$ and $\beta_r$ can be derived based on each observation to capture the nature variances and noises. 

For the Optimizer step, the objective is to determine the most cost-efficient size of SSD and RAM for the new machines that have 128 CPU cores. Instead of having a closed form as Equation~\ref{eq:obj} in Section~\ref{sec:tuning}, we use a Monte-Carlo simulation to estimate the objective function, i.e. the expected total cost of each configuration. 

We assume that the maximum number of running containers on a machine can be constrained by any of the three resources (CPU cores, SSD, and RAM). The cost of each configuration with different SSD and RAM sizes includes the penalty of idle CPU cores, SSD and RAM based on the unit cost of each resource and the extra penalty of running out of SSD or RAM. Running out of CPU is handled more gracefully in our system than running out of RAM or SSD. For a design with $S$ SSD and $R$ RAM, let $\alpha_s$ and $\alpha_r$ be the calibrated baseline usage for SSD and RAM respectively, we estimate the corresponding objective function by:

\begin{enumerate}
	\item Drawing random numbers $\beta_s$ and $\beta_r$ from the observational data;
	\item Calculating the maximum number of CPU cores that can be used, $c$, based on the inverse functions of $p$ and $q$ as following: 
	$$c = \min \{128, p^{-1}(S), q^{-1}(R)\}.$$
	\item Estimating the quantity of the idle resources. The number of idle CPU cores is:
	$$I_c = 128 - c.$$
	The amount of idle SSD is:
	$$I_s = S - p(c).$$
	The amount of idle RAM is:
	$$I_r = R - q(c).$$
	\item Estimating the total cost based on the unit price. If there is no idle SSD (RAM), the machine is stranded by SSD (RAM), adding an extra penalty for running out of SSD (RAM).  
\end{enumerate}

\begin{figure}[!t]
	\centering
	\begin{minipage}[b]{0.84\columnwidth}
		\includegraphics[width=\textwidth]{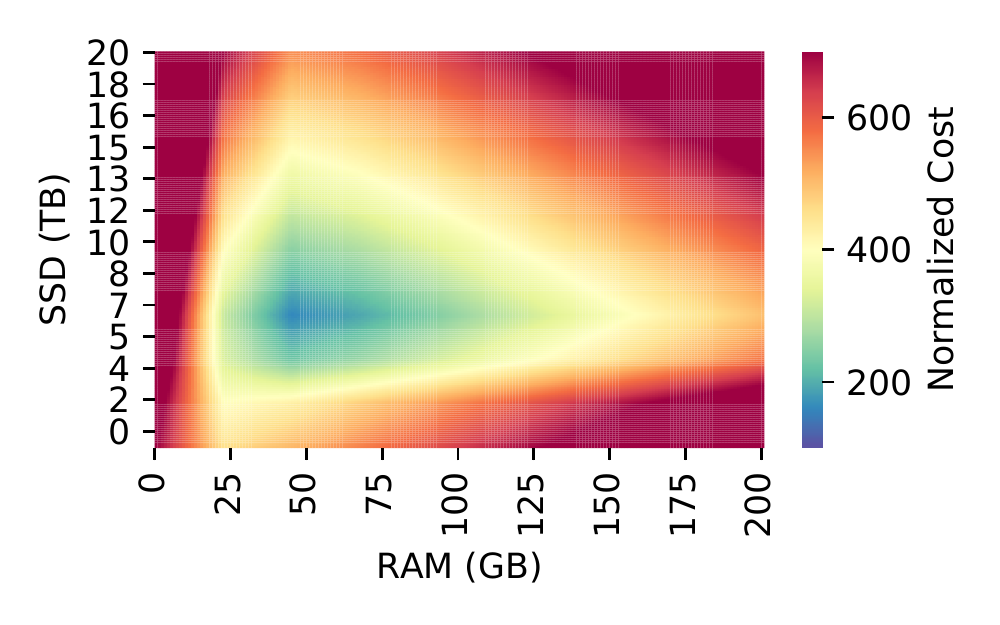}
	\end{minipage}%
	\vspace{-0.4cm}
	\caption{Expected cost with respect to different configurations}\label{fig:sku2}
\end{figure}

By repeating the above process 1000 times, we estimate the expected cost for each design configuration with different amounts of SSD and RAM (see Figure~\ref{fig:sku2}).
If the configuration is designed with insufficient SSD or RAM, the out-of-SSD or out-of-memory penalty dominates the cost. If the configuration is designed with too much SSD or RAM, the penalty of having idle resources increases. Therefore, we are looking for a ``sweet spot'' where the expected penalty based on the distribution of RAM and SSD usage per core is minimized (lower normalized cost in Figure~\ref{fig:sku2}). 

\subsection{Discussion}
The same methodology of \tone and \tthree is also applicable for optimizing other resources utilization, such as network bandwidth. The optimizer can take either a closed-form formulation and use commercial solvers, or use simple heuristics. In either case, given a predictor of the resulting performance (instead of building a complicated simulation platform), one avoids the need for experiments to deploy new configurations in the production cluster. The set of machine learning models precisely captures the system dynamics in the complex production environment, tailored to the customer workloads.

\section{\ttwo}\label{sec:pc}

Although \tone and \tthree cover a large number of applications, the performance impact for some configuration changes, such as changing a software configuration that affects the I/O speed or the introduction of a new feature to improve the processor performance, is still unpredictable. In this case, \ttwo is used.
With the introduction of machine-level metrics (see Figure~\ref{fig:metrics}), experiments can be done by the deploying experiments to groups of machines in production and conducting A/B testing at a smaller scale. 

In this section, we discuss the KEA applications that require experiments to deploy the configuration changes to a group of machines using the flighting tool and subsequent statistical analysis in Phase II (see Figure~\ref{fig:rcp}). 
For this group of applications, the key is the design of the experiments and the determination of performance metrics.

To have a fair comparison between the different groups of machines with different configurations, we hope to control the variables that can potentially affect the performance to the best effort, such as the hardware configurations, the time frame of data collection, even the physical location of the machines. To have statistical significance, we also want to have a relatively large sample size. In this paper, we summarize three possible experiment settings:
\begin{itemize}
	\item \textit{Ideal setting.} The ideal experiment setting is to have both the experiment group and control group from the same physical location, for example, choosing \textit{every other} machine in the same rack as the control/experiment group. 
	This setting is ideal as it ensures that the two groups of machines are receiving almost identical workloads throughout the experiment, and as they are physically located close to each other, they are often purchased at the same time, and storing data for similar customers. 
	However, sometimes this setting is not feasible, and further adjustments are required.
	\item \textit{Time-slicing setting.} The time-slicing setting is in general popular in A/B testing. For the same group of machines, this setting deploys consecutively the new and old configurations back-and-forth with a particular frequency, such as every five hours (instead of 24 hours to avoid day of week effects). The evaluation of different configurations is done by measuring the performance during different time intervals. However, this setting, even though it is popular in industry, has several limitations. In the production cluster, it is very difficult to frequently deploy new configurations, and workloads change in the different time intervals, therefore the selection of re-deployment interval becomes tricky. 
	\item \textit{Hybrid setting.} When the ideal or time-slicing settings are not feasible, we use the hybrid setting that deploys different configurations to different groups of machines. For this approach to be effective, we need to (i)~ensure that the groups of machines have similar characteristics, (ii)~use a sufficiently long time period, and (iii)~use metrics that are less sensitive to load/demand variation (e.g., normalized metrics).
\end{itemize}

In this section, we discuss two applications: selecting software configurations and power capping.

\subsection{Selecting Software Configurations}\label{sec:e2e3}

In this application, we achieve the ideal setting, i.e. selecting two rows (with approximately 700 machines each) and choose \textit{every other} machine in the same rack as the control/experiment group. 
We compare two different software configurations which represent using either SSD or HDD for the local temp store. \verb|SC1| puts local temp store on HDD and \verb|SC2| puts local temp store on SSD. 
The creation of the \verb|SC2| design was motivated by high local temp store write latency for \verb|SC1| caused by contention for I/O on the HDD. This write latency created a bottleneck for resource localization in our system.

The experiment was scheduled to run over five consecutive workdays. Table~\ref{table:e2e3} shows the performance impact using metrics that directly reflect the latency and throughput of the system.
The Total Data Read per day increased by 10.9\% while the average task latency decreased by 5.2\%, which is a very significant improvement. In all aspects of the performance of interest, the \verb|SC2| machines dominate and the result for Student's t-test shows that the changes are all significant.

\begin{table}[h]
	\caption{Performance metrics for different software configurations}\label{table:e2e3}
	\begin{tabular}{p{3cm}llcl}
		\toprule
		Name                     & \verb|SC1|      & \verb|SC2|      & \% Changes & t-value\\ \midrule
		Total Data Read (PB)    &1.38   & 1.53 & 10.9\%  & 40.4   \\
		Average Task Execution Time (s)       &24.1   & 22.9 & -5.2\%  &27.1   %
 \\\bottomrule
	\end{tabular}
\vspace{-0.2cm}
\end{table}

\subsection{Power Capping}

Compared to the experiment in the previous section, in this application, ideal experiment setting is not possible, i.e. dividing machines to the experiment and control groups in the same rack, as the power capping is at a higher level of control infrastructure and all machines in the same chassis have to be capped at the same level. Moreover, as we require multiple rounds of experiment to test the performance at different capping levels, the data will be collected for different time periods sequentially. Therefore, we use the hybrid setting in this application
and focus on the normalized metrics such as Bytes per CPU Time (ratio between Total Data Read and CPU time) and Bytes per Second (ratio between Total Data Read and task execution time), that are less sensitive to the workload level.

We start with the experiment by capping the machines to different provision levels and evaluate their performance. We also evaluate the performance impact for machines with a new feature at the processor level {\color{black}to accelerate processor and graphics performance}. 
For each round of the experiments with a particular level of capping, we collected data for four groups of machines for each SKU tested during the same time period to ensure that those groups of machines are receiving similar levels of workloads:
\begin{itemize}
	\item Group A with no capping and Feature off
	\item Group B with no capping and Feature on
	\item Group C with capping and Feature off
	\item Group D with capping and Feature on
\end{itemize}

We selected 120 machines for each group and capped the machines at 10\%, 15\%, 20\%, 25\% and 30\% below the original power provision level respectively. Each round of experiments ran for more than 24 hours.

\begin{figure}[!t]
	\centering
	\begin{minipage}[b]{0.5\columnwidth}
			\begin{tikzpicture}[thick,scale=0.6, every node/.style={transform shape}] 	
			\begin{axis}  
			[  
			ybar,
			width=7.2cm,
			height=5.5cm, %
			enlargelimits=0.15,%
			legend style={at={(0.5,1.25)}, %
				anchor=north,legend columns=-1},     
			ylabel={Perf Improvement (\%)}, %
			xlabel={Capping Level} ,
			ylabel near ticks, ylabel shift={-8pt},
			symbolic x coords={10\%, 15\%, 20\%, 25\%, 30\%},  
			xtick=data,  
			nodes near coords,  
			nodes near coords align={vertical},  
			label style={font=\huge},
			tick label style={font=\LARGE}
			]  
			\addplot coordinates {(10\%,5) (15\%,3.3) (20\%,1.2) (25\%,-2.6) (30\%,-7.8)}; %
			\addplot coordinates {(10\%,-0.9) (15\%,-0.4) (20\%,-2.2) (25\%,-4.8) (30\%,-10.9)};  
			\legend{Feature + Capping, Capping}  
			\end{axis}   
			\node[above,font=\huge\bfseries,align=center] at (current bounding box.north) {Bytes per CPU Time};
			\end{tikzpicture}  
	\end{minipage}%
	\begin{minipage}[b]{0.5\columnwidth}
		\begin{tikzpicture}[thick,scale=0.6, every node/.style={transform shape}] 		
		\begin{axis}  
		[  
		ybar,
		width=7.2cm,
		height=5.5cm, %
		enlargelimits=0.15,%
		legend style={at={(0.5,1.25)}, %
			anchor=north,legend columns=-1},     
		ylabel={Perf Improvement (\%)}, %
		xlabel={Capping Level} ,
		ylabel near ticks, ylabel shift={-8pt},
		symbolic x coords={10\%, 15\%, 20\%, 25\%, 30\%},  
		xtick=data,  
		nodes near coords,  
		nodes near coords align={vertical},  
		label style={font=\huge},
		tick label style={font=\LARGE} 
		]  
		\addplot coordinates {(10\%,5.2) (15\%,3.7) (20\%,1.9) (25\%,-1.3) (30\%,-7.8)}; %
		\addplot coordinates {(10\%,-0.4) (15\%,-0.6) (20\%,-1.4) (25\%,-4.4) (30\%,-10.4)};  
		\legend{Feature + Capping, Capping}  
		\end{axis}   
		\node[above,font=\huge\bfseries] at (current bounding box.north) {Bytes per Second};
		\end{tikzpicture}  
\end{minipage}%
	\vspace{-0.2cm}
	\caption{Performance impact due to power capping}\label{fig:pc}
	\vspace{-0.4cm}
\end{figure}
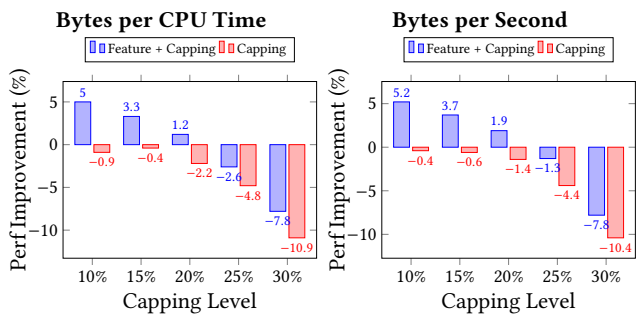

Figure~\ref{fig:pc} shows the performance impact on the two metrics due to different power capping limits for machines with a particular SKU with/without Feature enabled. The y-axis indicates the performance change benchmarked to the baseline, i.e. Group A with no capping and Feature off. With 10\% capping, with Feature enabled (blue bars), we can improve Bytes per CPU Time by 5\%. Without Feature enabled (orange bars), the same capping results in the performance degrading by 1\%. We can see that at higher power capping levels, the impact becomes more significant. In all cases, having Feature enabled improves the performance. 

Similar experiments were also conducted for other SKUs in different clusters to determine the optimal power provisioning limit. We have enabled a relatively conservative capping level which is still much lower than the original level and leads to approximately 10MW reduction in provisioned power that can be harvested to add more machines in the data center.

\subsection{Discussion}

The analysis for \ttwo is feasible because of the introduction of machine-level metrics that reflect the performance of the machines when running with a large amount of production traffic. It is impossible to isolate the impacts of configuration changes in the job-level metrics, as 
we cannot control for each job to be executed only in the experiment group or the control group (see Figure~\ref{fig:metrics}). 
We collect data for a relatively long time period to ensure that the machines received a sufficiently large amount of work and the performance is relatively stable for the evaluation of statistical tests.

The KEA infrastructure is used to evaluate many other features of the system and has become a standardized pipeline that leads to significant performance improvement.

\section{Related Work}
\label{sec:related}

Configuration tuning is vital for the performance of big data applications. Existing research focused on the tuning at the application level, leveraging experiment-based methods such as optimization~\cite{herodotou2011profiling,hua2018hadoop}, hill climbing~\cite{kamal2015performance,li2014mronline}, genetic algorithms~\cite{liao2013gunther}, Bayesian optimization~\cite{alipourfard2017cherrypick}, and other heuristic algorithms~\cite{zheng2007automatic,genkin2016automatic}. Similar methods are used for tuning DBMS instances on small-scale settings~\cite{van2017automatic,pavlo2019external,johannesBW}.

CherryPick~\cite{alipourfard2017cherrypick} used Bayesian Optimization~\cite{mockus2012bayesian,brochu2010tutorial,snoek2012practical} to pick the optimal configuration for Spark and Hadoop applications, such as VM instance type and cluster size. The model aims at minimizing the user cost, calculated as the product of running time and the price per unit with AWS EC2 services subject to a maximum job runtime. 
As suggested in~\cite{gardner2014bayesian}, the Expected Improvement (EI) algorithm was modified to incorporate this extra constraint by scaling the EI value by the probability of satisfying the running time constraint.

H-Tune~\cite{hua2018hadoop} is a learning-based configuration tuning approach for map-reduce applications. A two-level regression model was trained to predict the execution time at the phase and job level. 
Based on the prediction models, a genetic algorithm~\cite{mitchell1998introduction} was used to pick the optimal configuration. 

MRONLINE~\cite{li2014mronline} is an on-line tuning approach for map-reduce jobs running on YARN-based clusters. 
The tuner used a gray-box based hill climbing algorithm that consists of two steps, a global search phase to find a promising area and a local search phase to pick the best configuration. 

Gunther~\cite{liao2013gunther} is a search-based tuning algorithm for map-reduce jobs. A genetic algorithm~\cite{mitchell1998introduction} was used to pick the optimal configuration. The method was tested on two clusters with 16 slave nodes and the results show that with a small number of trials (less than 30), the method is able to obtain a new-optimal solution. 

Starfish~\cite{herodotou2011starfish} is a cost-based approach for optimizing map-reduce job performance in an offline setting using a ``what-if engine'' trained on job profile data.
Similarly, Selecta~\cite{klimovic2018selecta} proposed methods for auto-tuning cloud configurations for running spark jobs from the perspective of a cloud user.
These work focus on user pricing models where in KEA, we are approaching it from a cloud operator perspective resulting in a entirely different optimization problem. 

Zheng et al.~\cite{zheng2007automatic} proposed a software infrastructure to tune the configuration files in Internet services. One of their key contributions of this method is the notion of dependency graph between the tuned parameters, based on which an optimal searching order for parameters can be derived to improve efficiency.
Similarly, by measuring the performance of a target system under different configurations, \cite{kanellis2020too} developed a learning-based technique to identify the most important parameters to tune for achieving better performance in database systems. 

Note that all the above related work focuses on small-scale settings and relies on repetitive configuration deployment, which is not amiable (only used as last resort) in large-scale production environments such as ours.

{\color{black}ML and optimization techniques have also been used for resource allocation purposes to improve workload balance and reduce execution bottlenecks~\cite{adya2016slicer,albrecht2013janus,bianchini2020toward,sharov2015take,bortnikov2012predicting, hadary2020protean}. These works leverage information on resource utilization and (predicted) workload characteristics. In contrast, KEA focuses on tuning static configurations to affect scheduling, avoiding the need for dynamic real-time control components that introduce additional parameters and overheads.
}

\section{Conclusion}\label{sec:conclude}

In this paper, we propose KEA, a data-driven system for the configuration tuning of an exabyte-scale data infrastructure. 
Built upon an innovative way of conceptualization of the system components, KEA captures the
essence of our cluster dynamic behavior in a set of machine learning (ML) models and develops three tuning modes for different application scenarios (such as YARN configurations, hardware and data center design, and software investments). 
Evaluated in the production system, we demonstrate that KEA is able to continuously improve the performance step-by-step and introduce tens of millions of dollars saving per year for the company. Configuration tuning is becoming one of the biggest challenges for large scale systems, which might easily have hundreds even thousands of parameters. We believe that the KEA architecture can be extended for other configurations to further reduce the operational cost.

\bibliographystyle{ACM-Reference-Format}
\bibliography{sample-base}

\end{document}